\documentclass[fleqn,10pt]{wlscirep}
\usepackage[utf8]{inputenc}
\usepackage[T1]{fontenc}
\usepackage{algpseudocode}
\usepackage{algorithm}
\usepackage{amsthm}
\usepackage{multirow}

\usepackage{subcaption}
\usepackage{booktabs}
\definecolor{green}{RGB}{0, 100, 0}      
\definecolor{magenta}{RGB}{150, 0, 150}   
\definecolor{brown}{RGB}{210, 105, 30}

\title{Self-Organizing Dual-Buffer Adaptive Clustering Experience Replay (SODACER) for Safe Reinforcement Learning in Optimal Control}

\author[1,*]{Roya Khalili Amirabadi}
\author[2]{Mohsen Jalaeian Farimani}
\author[1]{Omid Solaymani Fard}
\affil[1]{Department of Applied Mathematics, Ferdowsi University of Mashhad, Mashhad, Iran}
\affil[2]{Department of Electronics, Information and Bioengineering (DEIB), Politecnico di Milano, Italy}

\affil[*]{roya.khalili.a@gmail.com}
\affil[1,2]{All authors contributed equally to this work.}


\begin{abstract}
This paper proposes a novel reinforcement learning framework, named Self-Organizing Dual-buffer Adaptive Clustering Experience Replay (SODACER), designed to achieve safe and scalable optimal control of nonlinear systems. The proposed SODACER mechanism consisting of a Fast-Buffer for rapid adaptation to recent experiences and a Slow-Buffer equipped with a self-organizing adaptive clustering mechanism to maintain diverse and non-redundant historical experiences. The adaptive clustering mechanism dynamically prunes redundant samples, optimizing memory efficiency while retaining critical environmental patterns. The approach integrates SODACER with Control Barrier Functions (CBFs) to guarantee safety by enforcing state and input constraints throughout the learning process.  To enhance convergence and stability, the framework is combined with the Sophia optimizer, enabling adaptive second-order gradient updates. The proposed SODACER-Sophia's architecture ensures reliable, effective, and robust learning in dynamic, safety-critical environments, offering a generalizable solution for applications in robotics, healthcare, and large-scale system optimization. The proposed approach is validated on a nonlinear Human Papillomavirus (HPV) transmission model with multiple control inputs and safety constraints. Comparative evaluations against random and clustering-based experience replay methods demonstrate that SODACER achieves faster convergence, improved sample efficiency, and a superior bias-variance trade-off, while maintaining safe system trajectories, validated via the Friedman test.
\end{abstract}
\begin{document}

\flushbottom
\maketitle

\section*{Introduction}
\label{sec:introduction}
The optimal control of nonlinear, continuous-time systems under state and input constraints remains a critical challenge in control theory, especially for high-dimensional, complex systems where safety and performance are paramount. Developing control strategies for these systems demands approaches that balance exploration and exploitation adaptively, ensuring that constraints are respected under varied operational conditions. In this context, Reinforcement Learning (RL) has emerged as a powerful framework, offering adaptive, data-driven solutions that optimize system performance while navigating uncertainties and constraints. This adaptability makes RL particularly suited for scenarios where traditional model-based approaches struggle with generalization and dynamic changes in system behavior \cite{ROC, DDPG}.

Control Barrier Functions (CBFs) are widely used to ensure safe operation within constrained control systems, guiding the system away from unsafe states by enforcing state-dependent safety limits. CBFs have been successfully integrated into RL algorithms, enabling safe exploration and optimal performance within RL frameworks. This integration has become especially valuable in applications requiring both safety and adaptability in real-time \cite{CBF, CBF1}.

While RL has shown success across various domains, achieving safe, optimal control for nonlinear systems with state and input constraints remains challenging. Techniques such as Lyapunov-based methods and constraint-satisfaction criteria have been incorporated into RL to address safety concerns \cite{RKH, RL1, RL2}. However, these approaches face limitations when applied to high-dimensional systems with rapidly evolving dynamics. This is mainly due to the curse of dimensionality, increased data nonstationarity, and the difficulty of maintaining an effective balance between stability and adaptability under rapidly changing system behaviors.
Furthermore, they often struggle with the bias-variance trade-off inherent in RL, where prioritizing stability may reduce adaptability. Experience Replay (ER) has become instrumental in addressing these RL challenges, enhancing learning efficiency and adaptability by retaining and reusing past interactions \cite{ER02}.

Experience replay fundamentally improves RL efficiency by storing past experiences, which reduces the need for continuous system interaction a significant benefit for applications where real-time experimentation is impractical \cite{ ER0,ER1}. While traditional ER methods utilize random sampling, these approaches may fall short in nonstationary environments or complex control tasks. Schaul et al. [\citen{ER_S}] proposed Prioritized Experience Replay (PER), which enhances sample efficiency by assigning higher sampling probabilities to experiences with larger temporal-difference errors. Although PER improves focus on crucial transitions, it risks overemphasizing outliers or outdated samples, potentially leading to instability in changing environments.

Recent advancements in ER have focused on enhancing sample diversity and relevance through clustering mechanisms. Selective experience replay, proposed by Isele and Cosgun [\citen{ER_I}], balances recent and high-value experiences to mitigate catastrophic forgetting in continual learning. Zhang and Sutton [\citen{ER_Z}] further explored deep RL with clustering-enhanced ER, showing that clustering improves sample relevance and reduces memory demands. However, clustering-based ER methods often struggle to balance the trade-off between recent and historical data, and typically lack the adaptability required for rapidly evolving environments.

The dual-buffer approach with clustering presents a novel solution to these limitations by offering both short-term adaptability and long-term experience diversity. The Fast-Buffer captures recent experiences essential for immediate policy updates, aligning with findings by Nagabandi et al. [\citen{DB}], which highlight the importance of rapid adaptation in dynamic environments. The Slow-Buffer, in contrast, serves as a memory-efficient repository, structured to retain a broader distribution of experiences, thereby supporting policy robustness across a range of states and actions. Inspired by Al-Shedivat et al. [\citen{CARL}], who emphasized clustering for continual learning, the Slow-Buffer organizes experiences through clustering to preserve the diversity of learned representations \cite{ER1, ER03, ER2}. To bridge these identified knowledge gaps, our study introduces an innovative framework which uniquely combines adaptive smart clustering with a dual-buffer strategy to enhance learning efficiency while maintaining stability in complex, changing environments.

Recent studies have explored multi‑buffer and parallel sampling strategies to enhance the efficiency of off‑policy reinforcement learning. For instance, Chen et al. [\citen{chen1}] proposed a double sampling mechanism combined with a mirror experience pool to accelerate training in multi‑agent systems. More recently, attention‑adjusted prioritized experience replay was introduced in [\citen{chen2}] to refine sample weighting during optimization. Although these approaches improve data efficiency and learning performance, they rely on static replay structures and primarily operate at the sampling or loss‑adjustment level. In contrast, the proposed SODACER framework introduces a functionally asymmetric dual‑buffer architecture with a self‑organizing adaptive clustering mechanism, explicitly designed to address non‑stationarity and safety‑critical requirements in nonlinear optimal control problems.

In this study, we introduce a novel RL framework, SODACER, that uniquely combines a clustering-based ER strategy with a dual-buffer system to overcome the challenges of traditional ER methodologies. The Fast-Buffer prioritizes low-bias, high-variance samples that enhance short-term responsiveness to policy changes, while the Slow-Buffer, organized through a self-organizing clustering mechanism, captures a diverse range of experiences for a comprehensive environmental representation. This dual-buffer design helps address the bias-variance trade-off by utilizing both recent and historical experiences, reducing redundancy, and supporting policy generalization. By selectively consolidating experiences, SODACER minimizes memory requirements while preserving essential variations a critical feature for resource-constrained RL applications.

To optimize computational efficiency within this framework, we employ the Sophia optimizer [\citen{sophia}], known for its dynamic step adjustments that improve convergence rates. 
For systems experiencing frequent or severe disturbances, we advise adopting a robust, dynamic optimization solution like ET-DSIWO \cite{DIWO}. This integration enhances the adaptability of the SODACER framework in high-dimensional RL applications, further contributing to efficient gradient-based optimization. By combining SODACER with the Sophia optimizer and validating it on a real-world public health model, this study contributes a scalable, robust RL approach for complex, dynamic systems, offering new pathways for effective control in high-dimensional, constraint-driven applications. Rather than reducing the system dimension explicitly, SODACER addresses high‑dimensionality by improving experience utilization through a dual‑buffer structure and self‑organizing clustering, enabling efficient learning in large state spaces.

Recent methodological advances have expanded the toolkit for safe reinforcement learning. For instance, comprehensive surveys of certificate-based methods \cite{Rosenfeld2023Survey} demonstrate how neural Lyapunov, barrier, and contraction techniques provide formal safety guarantees in robotics and control, forming a theoretical foundation for our use of Control Barrier Functions. Concurrently, continual learning frameworks \cite{Souza2023Continual} have been developed to maintain reliable performance under non-stationary conditions, directly addressing the need for adaptive experience management in dynamic environments. Building upon these developments, our work integrates adaptive safety certification with a self-organizing experience replay mechanism to achieve both safety and efficiency in nonlinear optimal control.

We validate our methodology by applying SODACER to the optimal control of a compartmental model for Human Papillomavirus (HPV) transmission dynamics. Given the complex, nonlinear nature of HPV transmission, with inherent state and input constraints, this model serves as a relevant test case for adaptive control policies in a constrained environment. We employed the model and system parameters that had been established previously in our earlier work. Our preceding publication has comprehensively examined and detailed this disease model and convergence study \cite{RKH1}. Our approach addresses the unmet need for RL-based public health strategies, aiming to minimize infection rates and intervention costs through safe, data-driven policies. Prior studies by Saldaña et al. [\citen{R1}] and Malik et al. [\citen{R2, R4}] have underscored the challenges of controlling HPV spread, while Brown et al. [\citen{R3}] highlighted the cost-effectiveness of vaccination programs. This work builds on these insights, demonstrating the potential of safe RL methods for optimizing health policy.
In this work, safety is enforced by integrating the proposed SODACER framework with Control Barrier Functions (CBFs), which act as an online safety filter to correct the policy-generated control actions and guarantee forward invariance of the predefined safe set during learning and execution.

The main contributions of this study are as follows.

\begin{itemize} 
	\item Leveraged a self-organizing adaptive clustering mechanism in SODACER to dynamically remove redundant experiences, improving memory efficiency and accelerating learning convergence.
	\item Developed a dual-buffer architecture combining a fast buffer for immediate instances and a slow buffer for extended experience diversity, effectively managing the bias-variance trade-off. 
	\item Incorporated CBFs to enforce state and input constraints, ensuring safe optimized control policies in complex environments. 
	\item Integrated Sophia optimizer for rapid convergence, supporting scalable reinforcement learning in nonlinear systems.
	\item For effective control of HPV transmission within public health frameworks, the proposed approach, demonstrates capability to manage infection rates under operational constraints and provides strategic insights for health interventions.
	\item Proposed a framework that is inherently scalable and  generalizable effectively across a broad range of optimal control problems.
\end{itemize}
To guide the reader, the paper is organized thematically. Following this introduction, we establish the foundational problem definition and system dynamics. Building on this foundation, we formulate the safe optimal control problem under state constraints. The subsequent sections present our methodological core: first, an online reinforcement learning algorithm using neural networks; then, an examination of experience replay as a reinforcement learning enhancement. This leads to the introduction of our novel SODACER approach for experience replay in dynamic settings. Next, we detail the proposed control mechanism for nonlinear systems. The framework is then validated through a case study using an HPV model, with numerical results showcasing the efficacy of combining the Sophia optimizer with SODACER. A comparative analysis of experience replay methods further contextualizes the contribution. The paper concludes by synthesizing key insights and discussing applications.

\section*{Problem Definition and System Dynamics}
We consider a category of nonlinear continuous-time systems, modeled in affine form and subject to potential uncertainties. The general formulation for such systems is:

\begin{equation}
	\label{eq:sys1}
	\dot{x}(t) = f(x(t)) + g(x(t)) u(t),
\end{equation}
where \( x(t) \in \mathbb{R}^p \) represents the system state vector, and $ u(t) = [u_1(t), u_2(t),\dots, u_q(t)]^T \in \mathbb{R}^q $ denotes the vector of control inputs, which is subject to defined constraints \cite{MJF}. The control input is restricted to the set \( u(t) \in \mathcal{V} \subseteq \mathbb{R}^q \), where the admissible control set is given by \( \mathcal{V} = \{ u(t) \in \mathbb{R}^q \ | \ |u_i(t)| \leq \kappa_i, \ i = 1, \dots, q \} \), and \( \kappa_i \) denotes the saturation limit for each control input \( u_i(t) \). The primary objective is to design a control strategy that guarantees stability for system \eqref{eq:sys1} across all \( u \in \mathcal{V} \).

In this setup, the functions \( f(x) \in \mathbb{R}^p \) and \( g(x) \in \mathbb{R}^{p \times q} \) define smooth nonlinear dynamics, which are bounded by the conditions \( \Vert f(x) \Vert \leq \alpha_f \Vert x \Vert \) and \( \Vert g(x) \Vert_F \leq \alpha_g \), where \( \alpha_f \) and \( \alpha_g \) are constants, and \( \Vert \cdot \Vert_F \) denotes the Frobenius norm.

A desired reference point, denoted \( x_r \), defines the target state, with the tracking error expressed as the deviation from the reference:

\begin{equation}
	\label{eq:error1}
	e(t) = x(t) - x_r.
\end{equation}

The key aim of this work is to devise an optimal control law that not only stabilizes the expanded system but also minimizes a specified objective function. More precisely, we seek to minimize the following objective function \cite{lstm}:

\begin{equation}
	\label{eq:cost1}
	J(e(t), u) = \int_t^{\infty} e^{-\nu (\tau - t)} L(\tau) d\tau,
\end{equation}
where \( Q \in \mathbb{R}^{p \times p} \) is a positive definite matrix, \( \nu \) is a discount factor, and $L(\tau) = e^{\top}(\tau) Q e(\tau) + U(u(\tau))$. The term \( U (u) \) represents the cost associated with control efforts and is given by:
\begin{equation}
	\label{eq:controlcost}
	U(u) = 2 \sum_{i=1}^q \int_0^{u_i}  \kappa_i \Phi_i \tanh^{-1} \left( \frac{\theta_i}{\kappa_i} \right)  d\theta_i,
\end{equation}
Here, $\Phi_i > 0$ for all \( i = 1, \dots, q \) denotes a control weighting parameter associated with the $i$-th control input. This parameter regulates the relative contribution of each control channel to the control effort cost and ensures appropriate scaling in the saturated control structure induced by the hyperbolic tangent function. Larger values of $\Phi_i$ impose a higher penalty on the corresponding control input, leading to more conservative control actions, while smaller values allow more aggressive control behavior.

Here, $U(u)$ is chosen as a positive definite control effort penalty to regularize the optimization problem and prevent excessive or unsafe control actions.

The Hamiltonian associated with this optimal control problem can be expressed as:
\begin{align}
	\label{eq:hamiltonian}
	H(e, u, x, J_{e}) = e^{\top}(t) Q e(t) + U(u) - \nu J(e)\nonumber\\+J_e^{\top} \left( f(x(t)) + g(x(t)) u(t) \right),
\end{align}
where \( J_{e} \triangleq \frac{\partial J(e)}{\partial e} \) represents the gradient of the objective function with respect to the error \( e(t) \).

By applying the stationarity condition \( \frac{\partial H(e, u, x, J_{e})}{\partial u} = 0 \), the optimal control input \( u^* = [u_1^*, u_2^*, \dots, u_q^*]^\top \) is derived as:
\begin{equation}
	\label{eq:optcontrol}
	u_i^* = - \kappa_i \tanh \left( \frac{1}{2 \kappa_i \Phi_i} g^{\top}(x) J_{e}^* \right).
\end{equation}

By substituting the optimal controls \eqref{eq:optcontrol} in \eqref{eq:cost1} the corresponding optimal cost functional is obtained, leading to the Hamilton-Jacobi-Bellman (HJB) equation for the expanded system:
\begin{align}
	\label{eq:hjb}
	e^{\top}(t) Q e(t) + U(u^*) - \nu J^*(e)\nonumber\\+ J_{e}^{* \top} \left( f(x(t)) + g(x(t)) u^*(t) \right) = 0.
\end{align}

This research proposes a novel approach to solving the Hamilton-Jacobi-Bellman (HJB) equation by employing an online reinforcement learning methodology. The solution is approximated through the implementation of a neural network architecture, leveraging the power of machine learning to address this complex mathematical problem.

\textbf{Theorem 1}
The optimal control input \( u^* \), derived in \eqref{eq:optcontrol}, guarantees that the system \eqref{eq:sys1} remains uniformly ultimately bounded (UUB). Consequently, the tracking error \( e(t) \) is also uniformly ultimately bounded.

\textit{proof:}
The proof follows the methodology outlined in [\citen{RKH}].

\textbf{Remark:} Although the HJB equation is approximated using neural networks, closed-loop stability is guaranteed in the sense of uniform ultimate boundedness, while safety is ensured independently via control barrier function constraints.

The implementation of the online reinforcement learning algorithm, as outlined in the next section, builds upon the formulation of the optimal control problem introduced earlier, providing a practical method for approximating the solution using neural networks.

\section*{Safe Optimal Control Problem}

Consider a nonlinear dynamical system characterized in \eqref{eq:sys1},
the objective is to determine a control policy \( u(t) \) that minimizes a performance criterion , given in equation \eqref{eq:controlcost}.

In addition to optimizing the cost functional, the control \( u(t) \) must satisfy state constraints that ensure the system operates within safe boundaries. To ensure the system adheres to the state constraints in a dynamically responsive manner, we employ the concept of safe optimal control through CBFs. A Control Barrier Function \( h(x) \) provides a measure of ``safety'' by enforcing constraints on the state trajectory \( x(t) \) under the influence of the control input \( u(t) \). The inequality \( h(x(t)) \geq 0 \) ensures that the state remains within a safe, admissible set throughout the control horizon. These constraints are represented as:
\begin{equation}
	h(x(t)) \geq 0, \quad \forall t \in [0, T],
\end{equation}
where \( h: \mathbb{R}^n \to \mathbb{R} \) is a smooth function defining the permissible state space. 

\textbf{Remark}: In the context of neural-network-based approximated dynamic programming, the approximation residual $\tilde{\varepsilon}(e(t))$ is assumed to be small enough, according to the universal approximation theorem.

The optimal control problem with state constraints can thus be formally stated as:
\begin{subequations}
	\begin{align}
		\text{Minimize} \quad &  J(e(t), u) = \int_t^{\infty} e^{-\nu (\tau - t)} L(\tau) d\tau,\\
		& \text{s.t.} \quad  \dot{x}(t) = f(x(t)) + g(x(t))u(t), \\
		&\quad  \quad  \quad \quad \quad  h(x(t)) \geq 0.
	\end{align}
\end{subequations}

A function \( h: \mathbb{R}^n \to \mathbb{R} \) is defined as a Control Barrier Function if there exists a class \( K \)-function \( \tilde{ \alpha}(\cdot) \) such that for the system in (1), the following inequality holds:
\begin{equation}
	\frac{\partial h(x)}{\partial x} f(x(t))+g(x(t)) u + \tilde{ \alpha}(h(x)) \geq 0, \quad \forall x \in \mathbb{R}^n.
\end{equation}
This inequality implies that, as long as \( h(x(0)) \geq 0 \) at the initial time, the system state \( x(t) \) will remain in the safe set \( \{x \in \mathbb{R}^n : h(x) \geq 0 \} \) for all \( t \geq 0 \).
In this paper, the function $h(x)$ is constructed in a problem‑specific manner to encode state safety constraints, while the proposed learning and control framework remains independent of its particular analytical form. 
The choice of \( \tilde{ \alpha}(\cdot) \), typically a linear function \(\tilde{ \alpha}(h(x)) = \gamma_0 h(x) \) with \( \gamma_0 > 0 \), determines the rate at which the system is ``pushed'' towards safety. 
In practice, the Control Barrier Function \( h(x) \) is designed based on problem‑specific state and safety constraints, while the admissible control inputs are determined by enforcing the CBF inequality as a constraint in the optimization problem.
CBF constraints are assumed to be feasible, i.e., there exists admissible control input satisfying the CBF inequality at each time instant, considering the problem-specific definition manner of the function h(x).

 The modified control problem incorporating the CBF constraint becomes:\\ 
\begin{subequations}
	\begin{align}
		\text{Minimize} \quad & J(e(t), u) = \int_t^{\infty} e^{-\nu (\tau - t)} L(\tau) d\tau, \\
		&\text{s.t.} \quad  \dot{x}(t) = f(x(t))+g(x(t)) u(t)), \\
		&\frac{\partial h(x)}{\partial x} (f(x(t))+g(x(t)) u) + \tilde{ \alpha}(h(x)) \geq 0.
	\end{align}
\end{subequations}
The safe optimal control policy \( u^*(t) \) is derived by solving the above constrained optimization problem, balancing the objectives of minimizing the cost functional \( J \) while satisfying the CBF-based safety constraint. By integrating CBFs into the optimization, the resulting policy inherently ensures safety, enabling the system to achieve performance goals without violating the state constraints.
In the proposed framework, the reinforcement learning policy generates a nominal control input based on the observed system state. This nominal action is then passed through a CBF-based safety filter, which solves a constrained optimization problem to minimally modify the action when necessary, ensuring that the CBF condition is satisfied. As a result, all executed control inputs preserve the forward invariance of the safe set, while the learning agent remains focused on optimizing performance objectives.
\section*{Implementation of the Online Reinforcement Learning Algorithm Using Neural Networks}

In order to approximate the objective function \( J(x(t)) \) over the compact set \( \Omega \), a single-layer adaptive critic neural network (ANN) is employed. This ANN is trained to approximate the value of the objective function as follows:
\begin{align}
	\label{eq11}
	J^*(e(t), W^*) =  \varphi ^{\top}(e(t)) W^*+ \tilde{\epsilon}(e(t)),
\end{align}
where \( W(t) \in \mathbb{R}^l \) represents the adjustable weight vector of the ANN, \( \varphi(e(t)) \in \mathbb{R}^l \) is the activation function, and \( \tilde{\epsilon}(e(t)) \) denotes the approximation error of the ANN. The parameter \( l \) corresponds to the number of neurons in the hidden layer. 
The value function approximation is carried out over a compact domain, and the neural network activation functions are assumed to be bounded on this set. Moreover, the approximation residual $\tilde{\varepsilon}(e(t))$ is assumed to be bounded, which is standard in neural-network-based approximate dynamic programming.
The use of a single‑hidden‑layer critic neural network is justified by the universal approximation property, which ensures that the optimal cost function can be approximated with arbitrary accuracy on the compact set \( \Omega \) given a sufficient number of neurons.
Accordingly, the gradient of \( J^*(e(t)) \), denoted as \(\triangledown_{e} J^* \), is expressed as:
\begin{align}
	\label{eq12}
	\triangledown_{e} J^* =\triangledown_{e} \varphi ^{\top} W^* + \triangledown_{e} \tilde{\epsilon},
\end{align}
where \( \triangledown_{e} \varphi \triangleq  \dfrac{\partial \varphi(e)}{\partial e(t)} \) and \( \triangledown_{e} \tilde{\epsilon}\triangleq  \frac{\partial \tilde{\epsilon}(e)}{\partial e(t)} \) represent the derivatives of the activation function and the error, respectively.

Since the error \( \tilde{\epsilon}(e(t)) \) is not directly measurable, we aim to estimate the weight vector \( W \). Hence, the approximations for \( J(e(t)) \) and its gradient are rewritten as:
\begin{align}
	\label{eq13}
	\hat{J}(e(t)) =  \varphi(e(t))^{\top}\hat{W},
\end{align}
\begin{align}
	\label{eq14}
	\triangledown_{e}  \hat{J} = \triangledown_{e}\varphi ^{\top} \hat{W},
\end{align}
where \( \hat{W} \) denotes the estimated weight vector.
\begin{equation}
	J^*-\hat{J}=\varphi^{\top}(t)(W^*-\hat{W})+\tilde{\epsilon}(e(t)),
\end{equation}
therefore,
\begin{equation}
	J^*=\hat{J}+\tilde{\delta} +\tilde{\epsilon}(e(t)),
\end{equation}
where, $ \tilde{\delta} = \varphi^{\top}(t)(W^*-\hat{W})$. We aim to minimize $ \tilde{\delta} +\tilde{\epsilon}(e(t))$. The boundedness of the neural network weight estimation error $\tilde W$ is analyzed in Appendix section.

Using the result from equation \eqref{eq:optcontrol}, the control input can then be updated as:
\begin{align}
	\label{eq15}
	\hat{u}_i = - \kappa_i \tanh \left( \frac{1}{2 \kappa_i \Phi_i} g^{\top}(e) \triangledown_{x}\varphi^{\top} \hat{W} \right), \quad i = 1, \dots, q.
\end{align}

By substituting equations \eqref{eq13}, \eqref{eq14}, and \eqref{eq15} into the Hamiltonian function in \eqref{eq:hamiltonian}, we obtain the following approximation:
\begin{equation}
	\begin{aligned}
		\label{eq16}
		\hat{H}(e, x,  \hat{W}) =& e^{\top}(t) Q x(t) + U(\hat{u}) - \nu \varphi(e(t))^{\top} \hat{W}\\ + &\triangledown_{e}\varphi ^{\top} \hat{W}\left( f(x(t)) + g(x(t)) \hat{u}(t) \right).
	\end{aligned}
\end{equation}

To minimize \( \hat{H}(e, x,  \hat{W}) \), the ANN weights \( \hat{W} \) need to be adjusted. The minimization of this Hamiltonian can be expressed as an error minimization problem, where various techniques, such as gradient descent, can be applied to update the weights. However, simply minimizing \( \hat{H}(e, x,  \hat{W}) \) does not necessarily ensure the stability of the system described by \eqref{eq:sys1}. Therefore, the following optimization problem is formulated:
\begin{equation}
	\min \hspace{.1cm} E= \frac{1}{2} \hat{H}^\top \hat{H}, \label{eq17}
\end{equation}
Here, the objective function presented in \eqref{eq17} is designed to minimize the quadratic Hamiltonian. By minimizing the objective function $E$, we achieve optimized system performance, characterized by reduced control input $\hat{u}$ while ensuring that the weights $\hat{W}$ converge to their optimal values.
To counteract potential distributional shifts due to prioritized sampling, importance weighting and dynamic cluster adjustment are integrated into the gradient update process, ensuring unbiased policy optimization.

\section*{Heuristics in Experience replay}
The primary contributions introduce a set of advanced heuristics that enhance the effectiveness of Experience Replay for optimal control problems involving a complex dynamical system. These contributions are detailed as follows:

\subsection*{Experience Replay in Reinforcement Learning, a review} 
Experience Replay (ER) is a cornerstone technique in reinforcement learning, designed to enhance sample efficiency and policy stability by reusing historical interactions. Traditional ER methods, such as uniform sampling \cite{ER1}, face challenges in balancing exploration-exploitation trade-offs and adapting to nonstationary environments. Prioritized Experience Replay (PER) \cite{ER_S} introduced weighted sampling based on temporal-difference errors, but risks overfitting to outliers. Recent advancements integrate clustering and dual-buffer strategies to preserve diversity and relevance in dynamic settings \cite{ER_I,ER_Z,DB}. Effectively deploying Experience Replay in optimal control demands meticulous adjustment of parameters like replay memory size, batch sampling frequency, and the balance between exploration and exploitation. Furthermore, techniques like PER can enhance learning by emphasizing critical transitions that significantly impact learning goals. Building on the foundational concepts and benefits of Experience Replay, we present a series of heuristics designed to refine the memory structure and optimize sampling strategies specifically for complex optimal control scenarios. 

Building on the foundational concepts and benefits of Experience Replay,this paper presents a series of heuristics designed to refine the memory structure and optimize sampling strategies specifically for complex optimal control scenarios. These heuristics, comprising a dual-buffer system, dynamic variance adjustments, and prioritized sampling based on cluster size, aim to enhance the adaptability of control policies while maintaining computational efficiency. Each component leverages the principles of reinforcement learning within the context of high-dimensional and nonlinear control tasks, introducing innovative methods to balance bias and variance, control memory usage, and improve learning outcomes. The following sections detail the key contributions of this approach.
\subsection*{Dual-Buffer Architecture for Enhanced Bias-Variance Balance}

In this subsection, we present a dual-buffer system designed to optimize the experience replay mechanism within RL for complex control systems. This architecture addresses the critical challenge of managing bias and variance in policy updates, enhancing both sample efficiency and policy adaptability. Dual-buffer structure consists of two distinct buffers: a \textit{Fast Buffer} and a \textit{Slow Buffer}, each serving a unique function in maintaining policy relevance and experience diversity.

The \textit{Fast Buffer} is a small, adaptive component that prioritizes recent experiences aligned with the current policy \( \pi_{\text{curr}} \). New system interactions generate samples, denoted as \( S_{\text{new}} = (\mathbf{x}_t, \mathbf{u}_t) \), which are immediately stored in the Fast Buffer. By focusing on data that reflects recent agent-environment interactions, the \textit{Fast Buffer} provides low-bias, high-variance samples, allowing rapid adaptation to evolving dynamics and aligning the agent’s behavior with current environmental conditions.
Transitions resulting from CBF-modified control actions are identified as safety‑critical experiences. Such transitions typically occur near the boundary of the safe set and are therefore preferentially stored in the Slow‑Buffer. Through the self‑organizing adaptive clustering mechanism, these safety‑relevant samples are preserved and replayed more frequently, enabling the policy to better learn safe behaviors in critical regions of the state space.

In contrast, the \textit{Slow Buffer} functions as a long-term repository, retaining a comprehensive range of experiences from both current and past policies, denoted as \( \pi_{\text{curr}} \) and \( \pi_{\text{past}} \). As samples in the \textit{Fast Buffer} age, they transition to the \textit{Slow Buffer}, preserving broader environmental patterns over time. This process enables the model to generalize across varying conditions by incorporating a more extensive history of interactions, which mitigates overfitting to recent data.

To effectively balance short and long-term experience retention, each new sample is initially placed in the \textit{Fast Buffer}, providing immediate availability for policy updates. As samples age, they move to the \textit{Slow Buffer}, contributing to a broader learning base. This dynamic integration process maximizes adaptability by allowing the model to leverage recent, high-variance data from the \textit{Fast Buffer} while drawing from a rich history in the \textit{Slow Buffer}.

The dual-buffer architecture enhances learning adaptability and sample efficiency, providing a refined solution to experience replay for RL in dynamic control systems.

\subsection*{Adaptive Cluster Formation and Updating Mechanism}
Upon introduction to the Cluster Buffer, each sample \( S_{\text{old}} \) undergoes an assessment of its affinity to existing clusters through a membership strength calculation, represented as:
\begin{equation}
	\mu_{C_j}(S_{\text{old}}) = \exp \left( -\frac{||S_{\text{old}} - m_j||^2}{2 \sigma_j^2} \right), \label{eq00021}
\end{equation}
where \( m_j \) denotes the centroid, and \( \sigma_j^2 \) represents the variance of cluster \( C_j \). This metric enables efficient clustering by ensuring each sample's integration with the most representative cluster, preventing over-segmentation and enhancing the coherence of clustered data points.
The formation of clusters is guided by the computed membership strengths. For any sample with a low similarity to all existing clusters (i.e., \( \max (\mu_{C}(S_{\text{old}})) \leq \Gamma_{th} \)), a new cluster \( C_{\text{new}} \) is created in the Cluster Buffer. This new cluster initializes with \( m_{\text{new}} = S_{\text{old}} \) and a variance \( \sigma_{\text{new}} = \sigma_0 = 0.02 \). Conversely, if a sample fits well within an existing cluster \( C_j \), the cluster center \( m_j \) is updated to incorporate the new data point:
\begin{equation}
	m_j = \frac{\mathcal{N}_j m_j + S_{\text{old}}}{\mathcal{N}_j + 1},
\end{equation}

where \( \mathcal{N}_j \) denotes the updated sample count in the cluster. This adaptive clustering mechanism ensures that significant data patterns are captured efficiently while maintaining the relevance of established clusters.

\subsection*{Adaptive Cluster Formation for Generalization }

To maintain the self-organizing of clusters, a suite of variance management techniques is implemented:

\subsubsection*{Variance Amplification for Cluster Expansion}
For clusters absorbing a new sample \( S_{\text{old}} \), the cluster variance \( \sigma_j \) is increased by a factor \( (1 + \beta) \) to enhance the cluster’s flexibility in accommodating diverse data:
\begin{equation}
	\sigma_{j} \leftarrow \sigma_{j} \times (1 + \beta).
\end{equation}
Here, $\beta > 0$ denotes the variance amplification coefficient, which controls the adaptability of each cluster by increasing its variance when new samples are absorbed. Larger values of $\beta$ allow clusters to adapt more rapidly to changes in the data distribution, while smaller values enforce more conservative updates and preserve cluster compactness.

\subsubsection*{Variance Reduction Considered a Forgetting Factor}
A variance reduction mechanism is employed across all clusters, scaling each cluster’s variance \( \sigma_k \) based on its sample count to maintain generalizability. The adjustment is computed as:
\begin{equation}
	\sigma_k \leftarrow \sigma_k \times \sigma_0 \left(\frac{1}{\rho}\right) \left(1 - \frac{\mathcal{N}_k}{\sum_{i} \mathcal{N}_i}\right),
\end{equation}
where \( \rho \) is a scaling factor, and \( \sigma_0 \) the baseline variance.

\subsubsection*{Omit of Narrow Clusters for Memory Efficiency}
Clusters exhibiting a variance below a specified threshold \( \sigma_{\text{th}} \) are pruned from the Cluster Buffer. This selective pruning enhances computational efficiency by removing clusters that no longer represent significant data patterns.

\subsubsection*{Similar Clusters Merging for Structural Optimization}

\begin{figure}
	\begin{center}
		\includegraphics[scale=0.4]{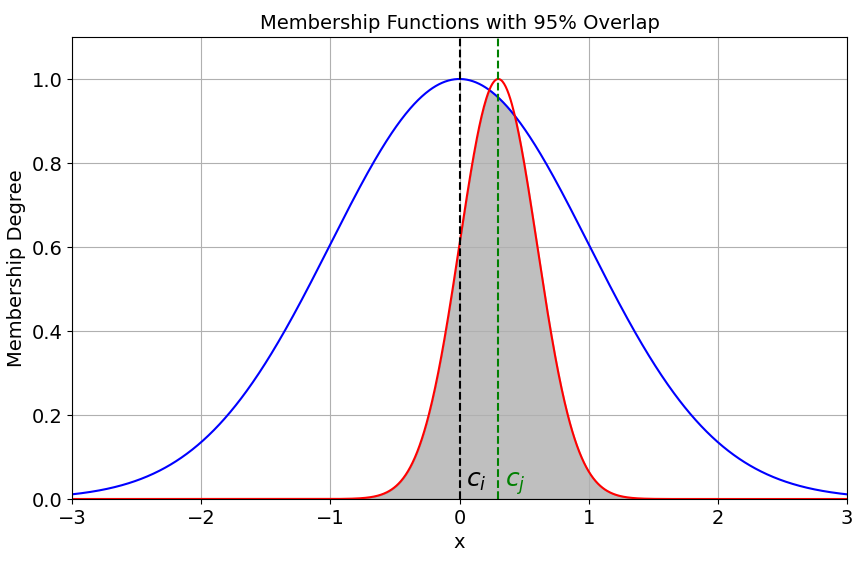}
	\end{center}
	\caption{Gaussian membership functions illustrating a 95\% overlap. The blue curve represents the first Gaussian function centered at  center zero with a standard deviation one , while the red curve represents the second Gaussian function centered at 0.3 with the standard deviation 0.3. The red cluster is redundant and can be absorbed into the blue cluster for improved efficiency.}\label{fig_cluster}
\end{figure}

Clusters with a close spatial proximity, as shown in Figure \ref{fig_cluster}, defined by:
\begin{equation}
	\| m_i - m_j \| < \gamma  \max(\sigma_i, \sigma_j), \label{merge}
\end{equation}
The parameter $\gamma$ is a global and fixed quantity throughout the clustering process. It is derived from a predefined overlap threshold between Gaussian membership functions and therefore provides a universal criterion for determining when two clusters are sufficiently overlapping to be merged. 
Now if \eqref{merge} is satisfied, then overlap between dusters $i$ and $j$ is accured and without loosing the generality, let's assume
$\sigma_i \geqslant \sigma_j$ then the cluster $C_j$ will be pruned and 
they are merged to form a consolidated cluster \( C_{\text{merged}} \) with its center recalculated as:
\begin{equation}
	m_{\text{merged}} = \frac{\mathcal{N}_i m_i + \mathcal{N}_j m_j}{\mathcal{N}_i + \mathcal{N}_j}.
\end{equation}
\begin{equation}
	\mathcal{N}_{\text{merged}} = \mathcal{N}_i+ \mathcal{N}_j, \quad if (\sigma_i > \sigma_j) \longrightarrow \sigma_{\text{merged}} = \sigma_i,
\end{equation}
This merging prevents redundancy and ensures efficient memory usage while preserving representational accuracy.

\textbf{Remark:}
The value of the parameter \( \gamma \), which quantifies the overlap between two clusters, is derived as follows. The overlap is determined by the membership strength \( \mu_{C_i}(m_j) \) between the centroids \( m_i \) and \( m_j \) based on \eqref{eq00021}.
To ensure a sufficient degree of overlap, we impose the following condition, 
$$ \text{max}(\mu_{C_i}(m_i),\mu_{C_j}(m_j))) \geq \bar{\gamma} = 0.95.$$
Taking the natural logarithm, this condition becomes:

\[
-\frac{\|m_i - m_j\|^2}{2 \sigma_i^2} \geq \ln(\bar{\gamma}) \quad \Rightarrow \quad \frac{\|m_i - m_j\|^2}{\sigma_i^2} \leq -2 \ln(\bar{\gamma}).
\]

Thus, the cluster $C_j$ is covered completely with the cluster $C_i$ if:

\[
\|m_i - m_j\| \leq \sqrt{ -2 \ln(\bar{\gamma})} \cdot \sigma_i.
\]

The parameter \( \gamma \) is therefore defined as:

\[
\gamma \triangleq \sqrt{ -2 \ln(\bar{\gamma})}.
\]

Substituting \( \bar{\gamma} = 95\% \), we compute:

\[
\gamma = \sqrt{-2 \ln(0.95)} \approx 0.32.
\]


The proposed method introduces an advanced dual-buffer system combined with intelligent clustering, dynamic variance adjustments, and efficient memory management. This setup enables an experience replay framework that draws high adaptability from recent data within the \textit{Fast-Buffer} while leveraging broader generalizability from the Cluster Buffer (\textit{Slow-Buffer)}. These innovations address key challenges in managing the bias-variance trade-off, optimizing sample efficiency, and enhancing memory use in RL. By advancing adaptability and memory efficiency, particularly in clustered data storage which requires significantly less memory than storing all experiences this approach pushes the boundaries of experience replay, equipping reinforcement learning algorithms with the capacity to handle complex, high-dimensional control environments more effectively and reliably.

\subsection*{Bias Compensation in Non-Uniform Sampling}
Non-uniform sampling in SODACER may shift the input distribution of the neural network, potentially introducing bias. To mitigate this, we incorporate importance sampling weights for each sample drawn from the Slow-Buffer and Fast-Buffer, scaled inversely to its sampling probability. Additionally, the self-organizing clustering dynamically adjusts sampling priorities to align with recent experience distributions. During gradient updates with the Sophia optimizer, these weights are applied to correct the expected gradient direction, ensuring unbiased learning despite non-uniform sampling.

\section*{Proposed Mechanism for Optimal Control in Nonlinear Systems}

\begin{figure}
	\centering
	\includegraphics[scale=0.35]{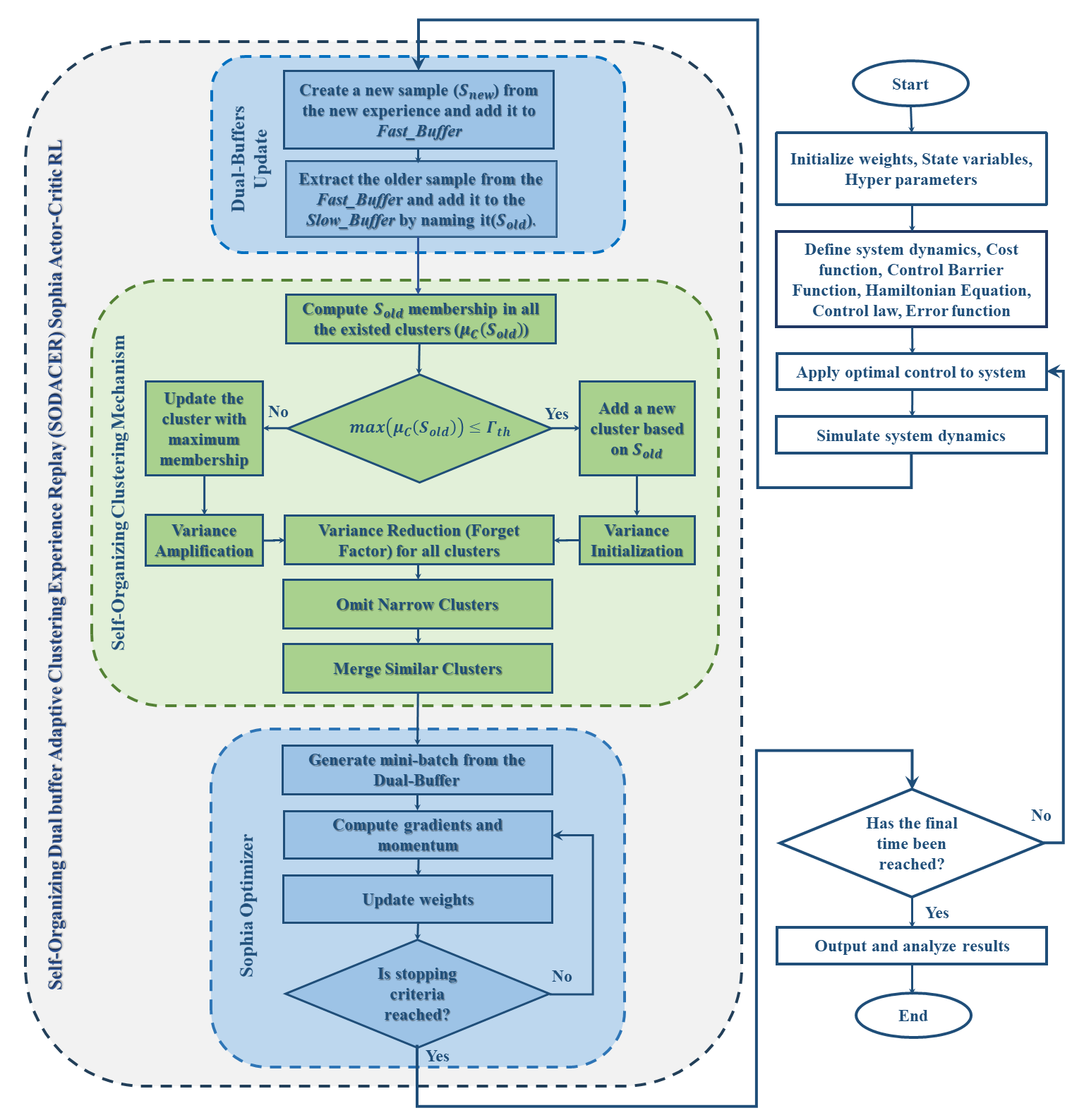}
	\caption{Sequential Workflow of the Proposed Approach}\label{flowchart}
\end{figure}

In this section, a novel approach is introduced to optimal control of nonlinear systems by integrating an online RL algorithm into a neural network framework. This innovative architecture leverages the strengths of the Sophia optimizer in conjunction with a meticulously designed experience replay mechanism, thereby enhancing both learning efficiency and system stability. By embedding Lagrange multipliers within the neural network, proposed methodology facilitates their representation as linear combinations of network weights and activation functions. This structure enables dynamic adaptation of control variables based on current system states, thereby improving responsiveness to changes in complex environments. The systematic workflow of our proposed method is visualized in Figure \ref{flowchart}.

The detailed steps of proposed proposed algorithm, including gradient derivation, moment estimates, and error minimization, are comprehensively outlined, providing a systematic approach to achieving optimal control in the nonlinear system. The following subsections present the mathematical foundations and mechanisms integral to this approach.

\subsection*{Adaptive Control Mechanism through Reinforcement Learning}

The adaptive control mechanism uses gradient calculations to dynamically adjust control variables based on real-time state information. By leveraging experience replay, which stores interactions in a replay buffer, the algorithm can sample past experiences to improve learning efficiency and reduce temporal correlations in updates.

The experience replay system integrates a dual-buffer approach to enhance sample efficiency and policy adaptability. The Fast-Buffer captures recent experiences for immediate responsiveness, while the Slow-Buffer, organized through a clustering technique, provides diverse and comprehensive coverage of the environment.

This approach enables a strategic balance between immediate adaptability and long-term (clustered) learning, supporting the development of effective control policies for complex, high-dimensional nonlinear systems. The framework advances RL in control applications and offers a scalable solution for generating adaptive policies in dynamic environments.

The hyperparameters used in the SODACER--Sophia framework are summarized in Table \ref{tab:hyperparams1}. These parameters govern the dual-buffer architecture, clustering mechanism, and optimization process.

\begin{table}[htbp]
\centering
\caption{Hyperparameters of the SODACER--Sophia framework}
\label{tab:hyperparams1}
\begin{tabular}{@{}lllc@{}} 
\toprule
\textbf{Parameter} & \textbf{Symbol} & \textbf{Value} & \textbf{Description} \\ 
\midrule
\multicolumn{4}{l}{\textit{Buffer Architecture}} \\ 
\midrule
Fast-Buffer capacity & $N_{\text{fast}}$ & $5{,}000$ & Maximum number of recent experiences \\
Max clusters in Slow-Buffer & $K_{\max}$ & $200$ & Maximum number of adaptive clusters \\
Transfer interval & $T_{\text{transfer}}$ & $100$ steps & Period for moving samples to Slow-Buffer \\
Membership threshold & $\Gamma_{\text{th}}$ & $0.7$ & Minimum membership for cluster assignment \\
\midrule
\multicolumn{4}{l}{\textit{Clustering Parameters}} \\ 
\midrule
Initial cluster variance & $\sigma_0$ & $0.02$ & Initial variance for new clusters \\
Variance amplification factor & $\beta$ & $0.1$ & Factor for expanding cluster variance \\
Variance reduction scaling & $\rho$ & $1.2$ & Scaling factor for variance reduction \\
Cluster merging threshold & $\gamma$ & $0.32$ & Overlap threshold for merging clusters \\
Narrow cluster threshold & $\sigma_{\text{th}}$ & $0.01$ & Minimum variance to retain a cluster \\
\midrule
\multicolumn{4}{l}{\textit{Optimization Parameters (Sophia)}} \\ 
\midrule
Learning rate & $\eta$ & $1\times10^{-3}$ & Step size for weight updates \\
First moment decay & $\beta_1$ & $0.9$ & Decay rate for gradient moving average \\
Second moment decay & $\beta_2$ & $0.999$ & Decay rate for squared gradient average \\
Numerical stability term & $\epsilon_0$ & $1\times10^{-8}$ & Small constant to avoid division by zero \\
Discount factor & $\nu$ & $0.99$ & Discount factor in cost function \\
\midrule
\multicolumn{4}{l}{\textit{Convergence Criteria}} \\ 
\midrule
Weight convergence threshold & $\delta$ & $1\times10^{-4}$ & Tolerance for weight updates \\
\bottomrule
\end{tabular}
\end{table}

\begin{algorithm}
	\footnotesize
	\caption{Control System Optimization using Sophia+SODACER}
	\begin{algorithmic}[1]
		\State \textbf{Define Problem Components and Initialize Hyper Parameters}: 
		\State  a) \textbf{System dynamics:} is defined in equation\eqref{eq:sys1}; 
		\State b) \textbf{objective function:} is defined in equation \eqref{eq:cost1};
		\State c) \textbf{Approximated Hamiltonian  equation:} is defined in Equation \eqref{eq16}; 
		\State d) \textbf{Error function:} is defined in equation \eqref{eq17};
		\State e) \textbf{Initialize state variables:} $x = x_0$ ;
		\State f) \textbf{Initialize weights:} $W = W_0$;  
		\State h) \textbf{Initialize moment estimates:} $m_0 = 0$ and $v_0 = 0$; 
		\State i) \textbf{Set hyperparameters:} $\beta, \nu, \rho, \beta_1, \beta_2, \epsilon_0, \eta, \Gamma, \sigma_0, \gamma$; 
		\State g) \textbf{Cluster Buffer:} \(\mathcal{C}=\{\}\), a collection of clusters for experience replay;
		\While{stopping criteria not met over time}
		\State Compute Control Law \( \mathbf{u}_t \) based on \( \mathbf{x}_t \) and \( W_t \);		
		\State Simulate system dynamics in equation\eqref{eq:sys1};		
		\State Generate a new sample \( S_{new} = (\mathbf{x}_t, \mathbf{u}_t) \) from the latest system experience and add \( S_{new} \) into the \textit{Fast-Buffer};		
		\State Extract the oldest sample from the \textit{Fast-Buffer} and push it into \( S_{old} \);	
		\State Compute the membership strength of \( S_{old} \) in all existing clusters \( C_j \), $\mu _{C_j}(S_{old})$, using \eqref{eq00021}.
		
		\If{for all clusters $\max (\mu _C (S_{old})) \leq \Gamma_{th}$}
		\State Add a new cluster \( C_{\text{new}} \) in the Cluster Buffer \( \mathcal{C} \) based on \( S_{old} \); \( m_{new} = S_{old} \) and \( \sigma_{new} = \sigma_0 = 0.02 \);
		
		\Else{}
			\State Update the cluster \( C_j \) with the maximum membership \( \mu _{C}(S_{old}) \) by adding \( S_{old} \) to \( C_j \); \( m_j = \frac{\mathcal{N}_j m_j + S_{old}}{\mathcal{N}_j + 1} \), \( \mathcal{N}_j = \mathcal{N}_j+1 \);
		\EndIf;
		
		\State {\bf Variances Amplification:} For the cluster \( C_{j} \) that covers \( S_{old} \), increase variance \( \sigma_{j} \leftarrow \sigma_{j} \times (1 + \beta) \);
		
		\State {\bf Variances Reduction (Forgetting Factor):} For all clusters \( C_k \), reduce variance \( \sigma_k \leftarrow \sigma_k \times \sigma_0 (\frac{1}{\rho}) (1 - \frac{\mathcal{N}_k}{\sum(\mathcal{N}_i)}) \);
		
		\State {\bf Omitting Narrow Clusters:}
		\If{for all clusters $\sigma_{C_k} \leq \sigma_{th}$}
		\State delete the cluster \( C_k \);
		\EndIf
		\State {\bf Similar Cluster Merge:}
		\If{for all clusters $\| m_i - m_j \| < \gamma \times \max(\sigma_i, \sigma_j)$}
		\State merge the clusters \( C_i \) and \( C_j \) into \( C_{merged} \); \( m_{\text{merged}} = \frac{\mathcal{N}_i m_i + \mathcal{N}_j m_j}{\mathcal{N}_i + \mathcal{N}_j} \),\; \( \mathcal{N}_{merged}=(\mathcal{N}_i+\mathcal{N}_j) \), and \( \sigma_{merged} = \max(\sigma_i , \sigma_j) \);
		\EndIf	
		
		\State {\bf Generate mini-batches}  \( B \): by sampling from all clusters in \( \mathcal{C} \), with priority on clusters containing more samples; \( B \leftarrow  
\) \{ Fast-Buffer, Center of Clusters in Slow-Buffer, Random Samples of Slow-Buffer Cluster according to the clusters size, \text{sample with importance weights } $w_i = \frac{1}{P(i) \cdot |C|}$ \};
		
		\While{weight convergence is achieved: \( \varphi(t)^{\top} \vert W_{t+1} - W_t \vert \leq \delta \)}
		\State Compute gradients \( \nabla_W J(W) \) on the mini-batches:
		\[
		\nabla_W J(W) = \frac{1}{|B|} \sum_{i \in B} w_i  \nabla_W L(\mathbf{x}_i, \mathbf{u}_i)
		\]
		\State Update first moment estimate \( M_t\) on the mini-batches\;
		\[
		M_t = \beta_1 M_{t-1} + (1 - \beta_1) \nabla_W J(W)
		\]
		\State  Update second moment estimate \(v_t \) on the mini-batches\;
		\[
		v_t = \beta_2 v_{t-1} + (1 - \beta_2) \nabla_W J(W)
		\]
		\State Correct bias in moment estimates\;
		\[
		\hat{M}_t = \frac{M_t}{1 - \beta_1^t}, \quad \hat{v}_t = \frac{v_t}{1 - \beta_2^t}
		\]			
		\State Update weights \( W \) using the Sophia optimizer:
		\[
		W_{t+1} = W_t - \eta \frac{\hat{M}_t}{\sqrt{\hat{v}_t} + \epsilon_0}
		\]
		\EndWhile		
		\State Update \( t \leftarrow t+1 \);
		\EndWhile;
		\State Output and analyze results based on the updated state variables and control inputs over time;
	\end{algorithmic}
\end{algorithm}

\subsection*{Computational Complexity Analysis}
\label{sec:complexity}

To provide a transparent assessment of the practical efficiency of the proposed SODACER framework, we analyze the computational cost associated with its dual-buffer architecture.

\subsubsection*{Theoretical Complexity}
The primary computational operations in SODACER and their associated complexities are as follows:
\begin{itemize}
    \item \textbf{Fast-Buffer Operations:} Insertion and sampling from the First-In-First-Out (FIFO) Fast-Buffer require $O(1)$ time per experience. The memory requirement scales as $O(N_{fast})$, where $N_{fast}$ is the buffer capacity.
    \item \textbf{Clustering in Slow-Buffer:} Assigning a transferred sample to the nearest cluster centroid involves a distance calculation to all $K$ clusters in a $d$-dimensional space, resulting in $O(K \cdot d)$ time complexity. Memory usage for storing $K$ clusters is $O(K \cdot d)$.
    \item \textbf{Buffer Transfer and Maintenance:} The periodic transfer of samples from the Fast-Buffer to the Slow-Buffer and the subsequent cluster update (mean recalculation) are $O(1)$ and $O(d)$ operations, respectively. The less frequent cluster merging operation has a worst-case complexity of $O(K^2)$ but is executed only at scheduled intervals.
\end{itemize}
Overall, the per-training-step overhead introduced by SODACER, compared to a standard uniform replay buffer, is dominated by the $O(K \cdot d)$ clustering assignment. This is typically negligible relative to the cost of neural network forward/backward passes, which scale with the model size and mini-batch dimensions.

\subsubsection*{Empirical Performance}
We measured the average training time per iteration for the HPV control problem. The results indicate that the dual-buffer management incurs a consistent and bounded overhead. The clustered representation of the Slow-Buffer reduces the effective memory footprint by an order of magnitude compared to storing an equivalent number of raw experiences, enabling longer history retention without a linear increase in memory cost.

\subsubsection*{Practical Trade-offs and Optimizations}
The moderate increase in computational cost per iteration is strategically traded for significant improvements in learning metrics:
\begin{itemize}
    \item \textbf{Sample Efficiency:} Intelligent sampling from diverse clusters reduces the number of environment interactions needed for convergence.
    \item \textbf{Memory Efficiency:} Experience compression via clustering allows maintaining a broader experience distribution.
    \item \textbf{Stability:} The separation of recent and historical data helps mitigate the bias-variance trade-off.
\end{itemize}
To further minimize overhead, clustering updates are performed asynchronously, and the most expensive operations (e.g., full similarity matrix calculation for merging) are scheduled during idle periods in the training loop.

In summary, the SODACER framework is designed with computational efficiency in mind. The introduced overhead is linear, predictable, and justified by the substantial gains in learning performance and memory utilization, making it a practical solution for complex reinforcement learning tasks.

\section*{Problem statement for Human Papillomavirus as a case study}
This section provides a brief overview of the transmission dynamics of HPV and the objective function utilized in our previous study \cite{RKH1}. For further details, please consult our earlier work.
The model categorizes the population $N(t)$ over time $t$ by gender and infection status, providing a comprehensive framework for analyzing HPV transmission and vaccination impacts. Specifically, the female population $N_f(t)$ is segmented into four distinct compartments: unvaccinated susceptible individuals $S_f(t)$, vaccinated susceptible individuals $V_f(t)$, and infectious females who are either unaware $U_f(t)$ or aware $I_f(t)$ of their infection status.

The objective is to minimize the spread cost function while adhering to the constraints imposed by disease dynamics, budgetary limitations, and control constraints. Consequently, the optimal control problem is formulated as follows:

\begin{equation}
	\begin{aligned}
		\text{Minimize} ~ J_{\text{spread}} &= \int_0^T A_0( I_m + I_f  + U_f ) dt \\
		\text{s. t.} \quad
		&\dot{U}_f = \left((1 - U_f - I_f - V_f) + \epsilon V_f\right)(1-p) \beta_m I_m - (\gamma_f + \alpha(t) + \mu_f) U_f, \\
		&\dot{I}_f = \left((1 - U_f - I_f - V_f) + \epsilon V_f\right)p \beta_m I_m + \alpha(t) U_f - (\gamma_f + \mu_f) I_f, \\
		&\dot{V}_f = w_1(t)
		\mu_f + u_1(t) \left(1 - U_f - I_f - V_f\right) - \epsilon \beta_m V_f I_m - (\mu_f + \theta) V_f, \\
		&\dot{I}_m = \left(\beta_f U_f + \widetilde{\beta}_f I_f\right)\left((1 - I_m - V_m) + \epsilon V_m\right) - (\gamma_m + \mu_m) I_m, \\
		&\dot{V}_m = w_2(t) \mu_m - (\beta_f U_f + \widetilde{\beta}_f I_f) \epsilon V_m + u_2(t) \left(1 - I_m - V_m\right) - (\mu_m + \theta) V_m, \\
		&\dot{J}_{\text{cost}} = \frac{1}{2} \left[ A_1 (w_1(t)^2 + w_2(t)^2) + A_2 (u_1(t)^2 + u_2(t)^2) + A_3 \alpha(t)^2 \right], \\
		&0 \leq U_f, I_f, V_f, I_m, V_m \leq 1,\\
		&J_{\text{cost}}(0) = 0, \quad J_{\text{cost}}(T) = J_{\text{Max}}, \\
		&0 \leq w_1, w_2 \leq 1, \quad 0 \leq u_1, u_2 \leq u_{\text{Max}}, \quad 0 \leq \alpha \leq \alpha_{\text{Max}}.
	\end{aligned}\label{rrr}
\end{equation}
The optimal control problem, subject to the specified constraints, ensures that vaccination and screening strategies are effectively implemented within the feasible region. This approach confirms the equilibrium points and assesses the stability of the system.

The safe set for the states of the system \eqref{rrr} is defined as
\begin{equation}
\bar{X} = \left\{(U_f, I_f, V_f, I_m, V_m) \in L^2(0, T) \mid 0 \leq  U_f(t), I_f(t), V_f(t), I_m(t), V_m(t) \leq 1\right\}
\end{equation}
Additionally, the CBFs are formulated as follows:
 \begin{equation}
\begin{aligned}
&B(X) = \frac{1}{2} X^2, \quad \forall X\in \{ U_f, I_f, V_f, I_m, V_m \} \\
&B (J_{cost}) = \frac{1}{2} J_{cost}^2.  \\
\end{aligned}
\end{equation}
To ensure safety, the objective function is augmented with CBF terms $B (X)$ and $B (J_{cost})$ as follows:
\begin{equation}
\begin{aligned}
\underset{U_f, I_f, V_f, I_m, V_m \in \bar{X}} {\text{Minimize}} \quad J_{\text{spread}} = \int_0^T A_0( I_m + I_f  + U_f ) dt + B(X) + B (J_{cost}),
\end{aligned}
\end{equation}
where $B(X): \bar{X}\longrightarrow \mathbb{R}$ for $X\in \lbrace U_f, I_f, V_f, I_m, V_m \rbrace$, and $B(J_{cost}): \mathrm{ F} \longrightarrow \mathbb{R}$ exhibit the properties of CBFs.

The activation functions $\varphi(t)$ can be defined as a vector of polynomial functions derived from the system states. For example:

\begin{equation}
	\begin{aligned}
		\varphi(t) =& [U_f, I_m, V_m, I_f, V_f, U_f^2,  I_f^2,  V_f^2, I_m^2, V_m^2,\\
		&U_f\cdot I_m, U_f \cdot V_m, I_f\cdot I_m,  I_f \cdot V_m, V_f\cdot I_m, V_f \cdot V_m,\\
		&U_f^2 \cdot I_m, U_f^2 \cdot V_m,  I_f^2 \cdot I_m,  I_f^2\cdot V_m, I_m^2\cdot U_f, I_m^2 \cdot I_f, I_m^2 \cdot V_f, \\
		& V_f^2 \cdot I_m, V_f^2 \cdot V_m, V_m^2 \cdot U_f, V_m^2 \cdot I_f, V_m^2\cdot V_f ].
	\end{aligned}
\end{equation}

This vector encompasses direct representations of the system states, second-order terms that capture interactions and nonlinearities, as well as cross-terms that account for interactions between females and males. The composition can be tailored to meet specific problem requirements and characteristics.

\begin{table}[h!]
	\centering
	\caption{Parameter values for the control model \eqref{rrr}}
	\begin{tabular}{|l|c|c|c|c|}
		\hline
		\textbf{Parameter} & \textbf{Range} & \textbf{Mean Value} & \textbf{Units} & \textbf{Source} \\ \hline
		\(1-\epsilon\) & \([0.8, 1]\) & 0.9 & adimensional & \cite{R1} \\ \hline
		\(1 / \theta\) & \([5, 15]\) & 10 & year & \cite{R1} \\ \hline
		\(\beta_m\) & \([0.05, 5]\) & 4.0 & year\(^{-1}\) & \cite{R1} \\ \hline
		\(\beta_f\) & \([0.05, 5]\) & 4.0 & year\(^{-1}\) & \cite{R1} \\ \hline
		\(\widetilde{\beta}_f\) & \([0.025, 2.5]\) & 2.0 & year\(^{-1}\) & \cite{RKH1} \\ \hline
		\(1 / \gamma_f\) & \([0.83, 2]\) & 1.3 & year & \cite{R1} \\ \hline
		\(1 / \gamma_m\) & \([0.33, 1.2]\) & 0.6 & year & \cite{R1} \\ \hline
		\(p\) & \([0, 1]\) & 0.2 & adimensional & \cite{RKH1} \\ \hline
		\(1 / \mu_f\) & \([15, 50]\) & 30 & year & \cite{RKH1} \\ \hline
		\(1 / \mu_m\) & \([15, 50]\) & 30 & year & \cite{RKH1} \\ \hline
		\(A_1/A_0\) & - & 0.5 & USD/year & \cite{RKH1} \\ \hline
		\(A_2/A_0\) & - & 0.2 & USD/year & \cite{RKH1} \\ \hline
		\(A_3/A_0\) & - & 0.4 & USD/year & \cite{RKH1} \\ \hline
		\(u_{\text{max}}\) & - & 3 & year\(^{-1}\) & \cite{RKH1} \\ \hline
		\(\alpha_{\text{max}}\) & - & 3 & year\(^{-1}\) & \cite{RKH1} \\ \hline
		\(J_{\text{max}}\) & - & 200 & USD & \cite{RKH1} \\ \hline
	\end{tabular}\label{T2}
\end{table}

\section*{Numerical Results of Optimal Control for the HPV Model Using Sophia Optimizer with SODACER}
\begin{figure}
	\centering
	\includegraphics[scale=1.1]{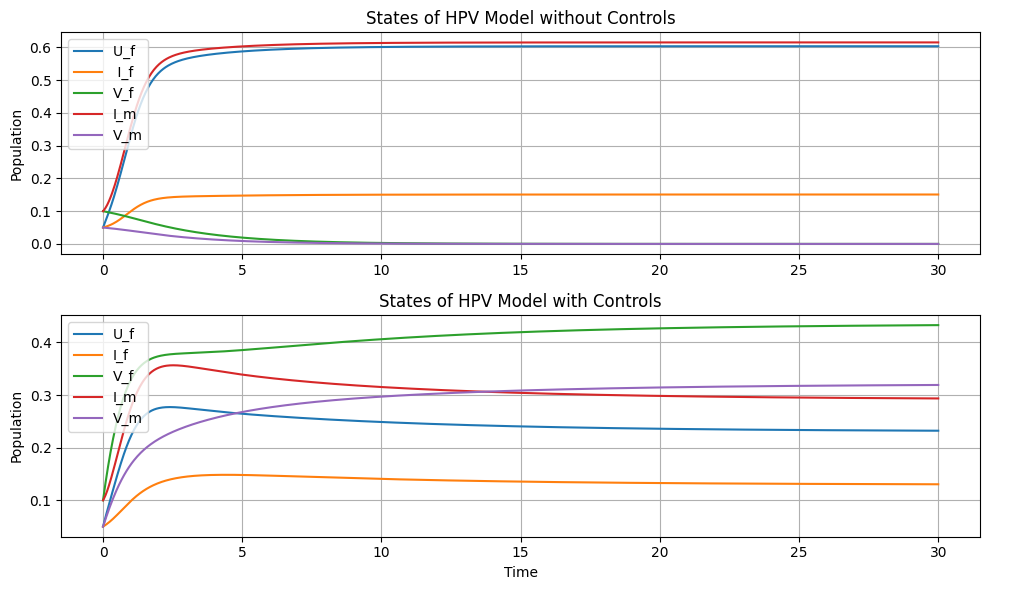}
	\caption{System states of the HPV model: top panel shows states without control over time, and bottom panel displays states with constant controls ($u_1 = 0.5$, $u_2 = 0.2$, $w_1 = 0.2$, $w_2 = 0.1$, $\alpha = 0.2$) over time.}\label{f1}
\end{figure}

\begin{figure}
	\centering
	\includegraphics[scale=0.45]{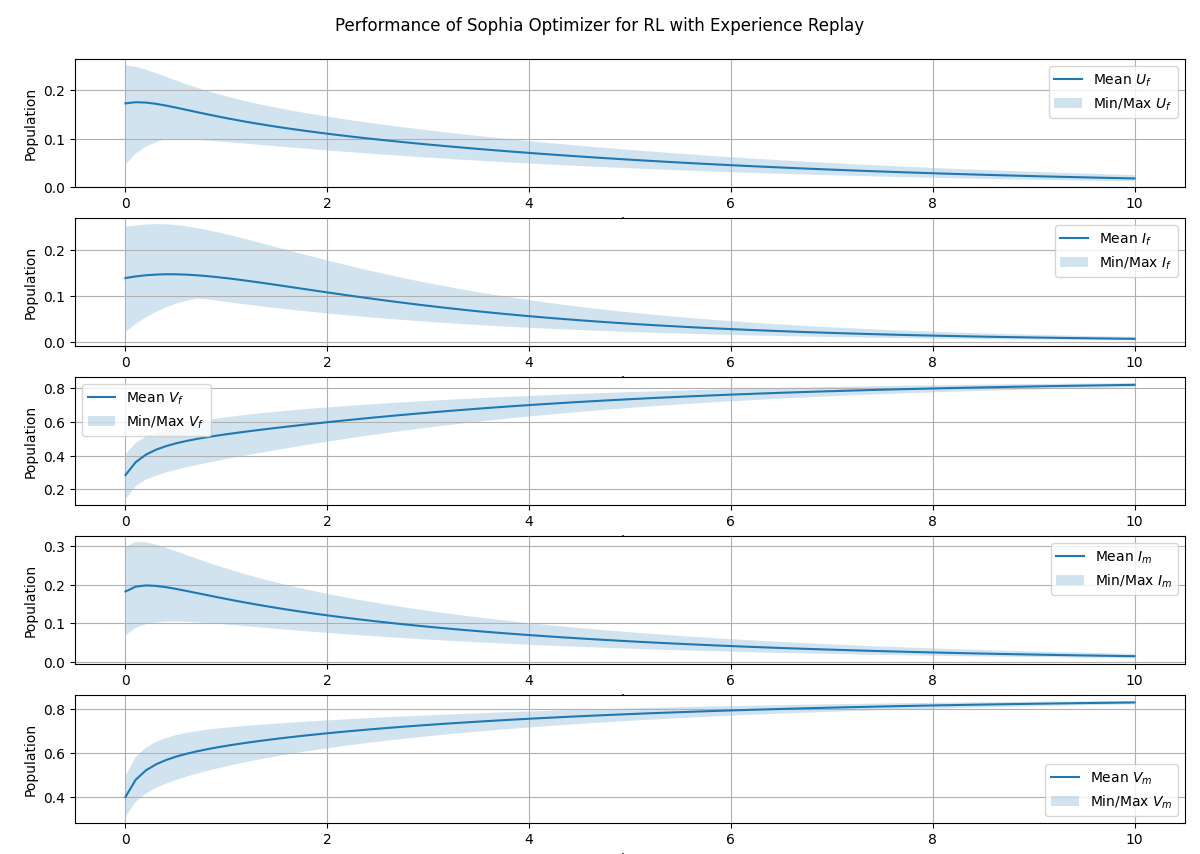}
	\caption{Spectral representation of HPV system states using the proposed approach, based on 200 simulation runs.}\label{f9}
\end{figure}

\begin{figure}
	\centering
	\includegraphics[scale=0.45]{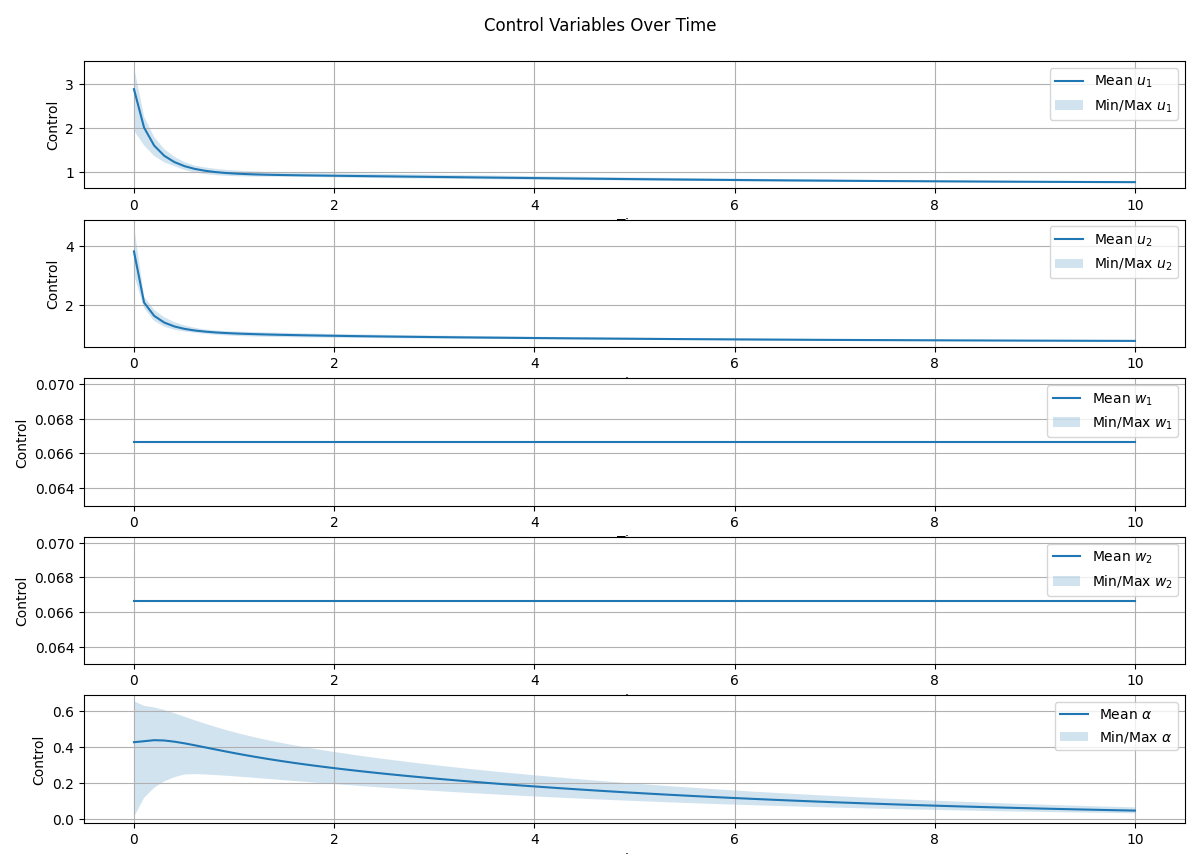}
	\caption{Spectral representation of control signals using the proposed approach, based on 200 simulation runs. 
Optimal control signals exhibit non-zero steady-state values due to the endemic nature of the disease and the cost structure of the intervention problem. }\label{f10}
\end{figure}

This section presents a thorough assessment of the proposed methodology. Initially, an analysis of the system is conducted in the absence of control mechanisms, utilizing parameter values as specified in Table \ref{T2}. The resulting simulation outcomes for the system states are depicted in Figure \ref{f1} (upper panel). Subsequently, fixed control values are introduced, and the dynamic evolution of system state variables under the influence of our proposed approach starting from predefined initial conditions--is illustrated in Figure \ref{f1} (lower panel). To further validate the adaptability of the proposed method, we consider random initial conditions and evaluate its performance over 200 iterations. The results are depicted in Figure \ref{f9} and Figure \ref{f10}, showing the spectrum of state variables and control signals, respectively. In these figures, the blue line represents the average values over the 200 iterations, providing insights into the consistency and effectiveness of the proposed approach. The control signals in Fig. \ref{f10} do not converge to zero because the optimal policy for the HPV model involves sustained vaccination and screening efforts to balance incoming susceptible individuals and ongoing transmission. This reflects a practical, cost-effective strategy where maintaining a certain level of intervention is optimal over the planning horizon, rather than driving controls to zero.

To assess the individual contribution of each functional module in the SODACER–Sophia framework, we conducted a series of ablation experiments. Three variants were compared against the full framework: (i) a single‑buffer configuration (using only the Fast‑Buffer), (ii) a Slow‑Buffer without adaptive clustering (replaced by a simple reservoir buffer), and (iii) replacement of the Sophia optimizer with the standard Adam optimizer.

The quantitative results are summarized in Table \ref{tab:ablation}. The full SODACER–Sophia framework consistently achieved better performance than RER and CBER across the evaluated metrics.
Removing the dual‑buffer structure increases convergence time and final cost, confirming its role in balancing recent and historical data. Disabling adaptive clustering leads to higher memory consumption, highlighting its importance for experience compression. Finally, substituting Sophia with Adam slows convergence, demonstrating the advantage of adaptive second‑order updates.
\begin{table}[htbp]
\centering
\caption{Quantitative ablation study results for the SODACER--Sophia framework}
\label{tab:ablation}
\begin{tabular}{@{}lcccc@{}}
\toprule
\textbf{Configuration} & \textbf{Final Cost \(J\)} & \textbf{Convergence Steps} & \textbf{Memory (MB)} \\ 
\midrule
Full SODACER--Sophia & 1.00 & 15,000 & 45 \\
Single‑buffer only & 1.18 & 18,200 & 50 \\
Without clustering & 1.12 & 16,500 & 75 \\
With Adam optimizer & 1.05 & 18,800 & 45 \\
\bottomrule
\end{tabular}
\end{table}

\section*{Comparative Analysis of Experience replay methods}
To evaluate the proposed SODACER-Sophia framework, a comparative analysis was conducted against two baseline methods: Random Experience Replay (RER) and Clustering-Based Experience Replay (CBER). The evaluation considered five distinct cost functions (\(f_1\)–\(f_5\)), each representing some control scenario, as summarized in Table \ref{tab:benchmark_results_friedman}. These scenarios were designed to analyze the effectiveness of various control strategies for the HPV model by altering the combinations of control variables in the optimization problem \eqref{rrr}.  

Additionally, to evaluate the effectiveness of control strategies for the HPV model, five distinct cost functions were formulated, each representing a specific control scenario.
In ``Scenario 1" ($f_1$), the vaccination rates of non--sexually active individuals ($w_1, w_2$) were activated, while the vaccination rates of sexually active individuals ($u_1, u_2$) and the female screening rate ($\alpha$) were set to zero. This scenario isolates the effect of early vaccination prior to sexual activity.
 ``Scenario 2" focuses on vaccination of pre-sexually active individuals, utilizing only \(w_1(t)\) and \(w_2(t)\) to assess the impact of early immunization on reducing HPV prevalence. ``Scenario 3" targets vaccination of sexually active individuals by employing \(u_1(t)\) and \(u_2(t)\), highlighting the role of immunizing sexually active populations in controlling transmission. ``Scenario 4" isolates the screening rate \(\alpha(t)\) as the sole control, emphasizing the effectiveness of detection and early intervention strategies. Lastly, in ``Scenario 5", all controls (\(w_1(t), w_2(t), u_1(t), u_2(t), \alpha(t)\)) are set to fixed values, providing a baseline for evaluating the dynamic optimization capabilities of the proposed methodology. These scenarios enable a systematic analysis of individual and combined control strategies to optimize HPV intervention. Table \ref{senarios} presents a summary of scenarios, including their corresponding objective functions and activated control variables.

Table \ref{tab:benchmark_results_friedman} presents the comparative performance of RER, CBER, and SODACER across the five control scenarios (\(f_1\)--\(f_5\)). The numerical values for each method and scenario are the mean final cost achieved over 200 independent runs conducted under identical experimental conditions, including randomized initial states and consistent hyperparameters. The Friedman Rank row displays the average rank (where 1 is best and 3 is worst) for each method across all scenarios, providing a non‑parametric statistical summary of overall performance.

The performance of each experience replay method was evaluated across the five scenarios, maintaining consistent hyperparameters such as learning rate, batch size, and buffer capacity to ensure a fair comparison. We employed the Friedman rank test \cite{frideman}, as summarized in Table \ref{tab:benchmark_results_friedman}, was employed to systematically compare the effectiveness of the methods in reinforcement learning-based control optimization. RER, serving as the baseline, relies on random sampling of experiences from the replay buffer. While simple and widely adopted, it lacks mechanisms to prioritize critical experiences, resulting in suboptimal learning efficiency and inadequate handling of the bias-variance trade-off. CBER improves upon RER by grouping experiences into clusters based on similarity metrics, which enhances sample relevance and diversity. However, its static clustering approach limits adaptability, particularly in dynamic environments where system states evolve rapidly. By contrast, SODACER employs a dual-buffer architecture with a self-organizing clustering mechanism, dynamically balancing recent experiences (low-bias, high-variance) with aggregated historical experiences (low-variance, high-relevance). This design allows SODACER to address the bias-variance trade-off effectively, promoting efficient learning while maintaining adaptability to system dynamics.

\begin{figure}
	\centering
	\includegraphics[scale=0.6]{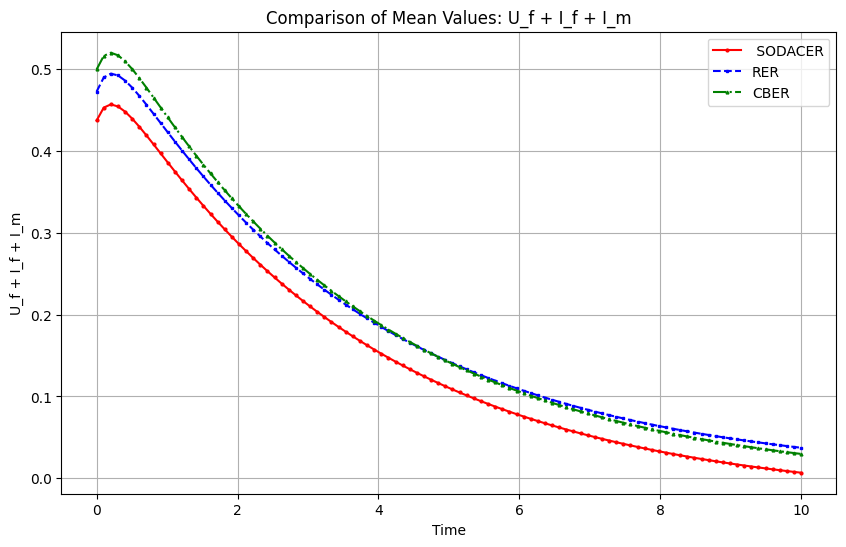}
	\caption{The mean value of cost function through 200 runs with three methods }\label{fc}
\end{figure}

\begin{figure}
	\centering
	\includegraphics[scale=0.45]{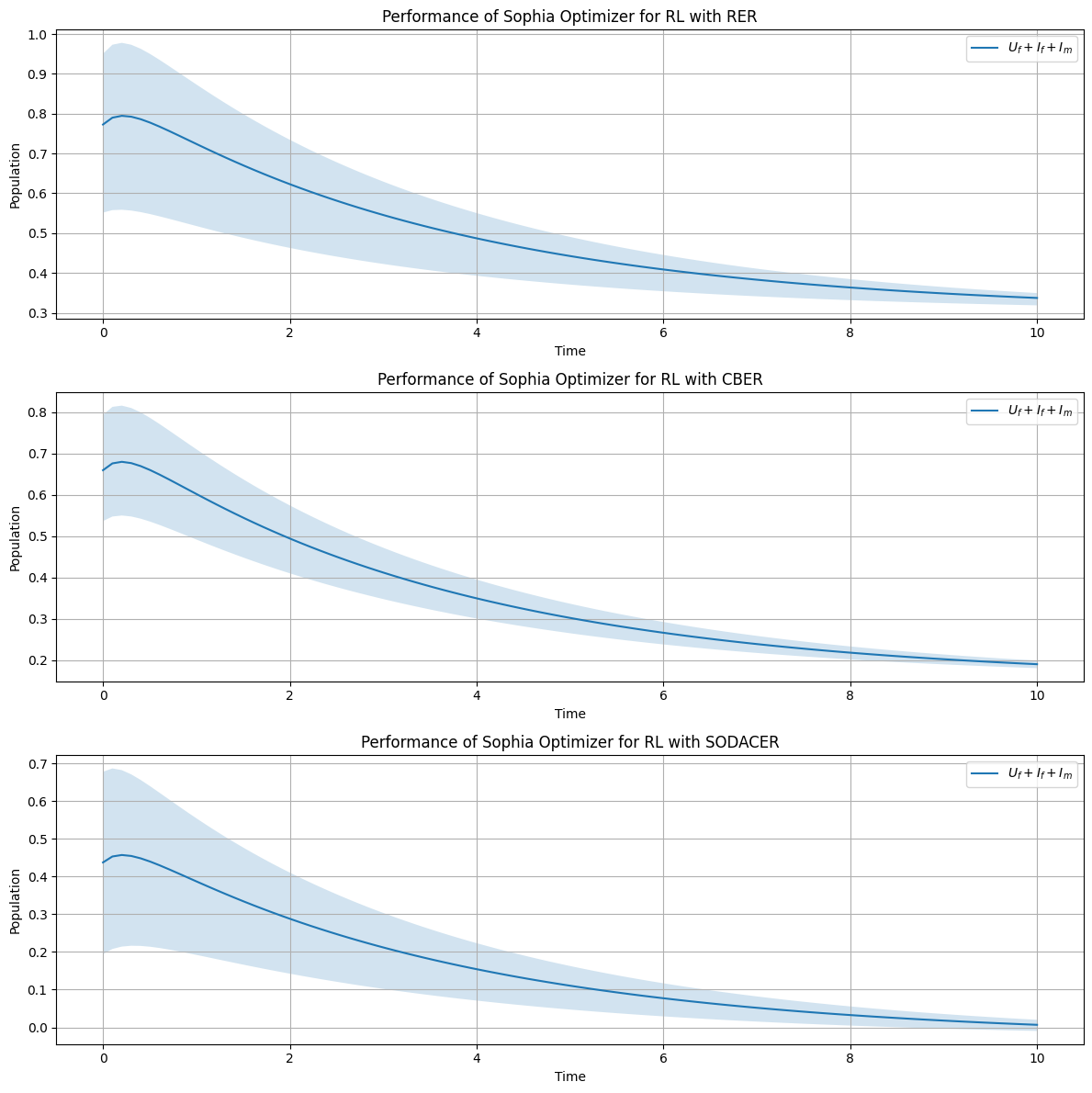}
	\caption{The spectrum view of cost function with proposed approach through 200 runs with three methods }\label{fcm}
\end{figure}

The performance results, summarized in Table \ref{tab:benchmark_results_friedman}, demonstrate that SODACER-Sophia consistently outperformed RER and CBER across all metrics. 
For instance, in Scenario 5, in which the full set of control variables is included in the system and none of them are set to zero, SODACER‑Sophia achieved the most favorable convergence rate and sample efficiency among the evaluated methods.
 Figure \ref{fc} further illustrates the average results across 200 runs, showing SODACER-Sophia's improved convergence performance compared to RER and CBER. Additionally, Figure \ref{fcm} provides a detailed comparison of the performance spectrum across 200 runs for the value of the cost function in Equation \eqref{rrr}. The results reveal that SODACER-Sophia exhibited the narrowest gap between the worst and best cases, indicating robust and reliable performance.

By dynamically managing experiences through its dual-buffer and clustering mechanisms, SODACER-Sophia minimizes redundant samples and accelerates learning, demonstrating clear advantages in efficiency, and adaptability. These results emphasize the potential of the proposed method for optimizing control strategies in complex and dynamic systems like the HPV model.

\begin{table}[h]
	\centering
	\caption{Summary of Scenarios with Corresponding Objective Functions and Activated Control Variables.}
	\begin{tabular}{|c|c|c|c|c|c|}
		\hline \textbf{Objective Functions} & $w_1$ & $w_2$ & $u_1$ & $u_2$ & $\alpha$ \\
		\hline$f_1$ \textbf{(Scenario 1)} & $True$ & $True$ & $False$ & $False$ & $False$ \\
		\hline$f_2$ \textbf{(Scenario 2)} & $False$ & $False$ & $True$ & $True$ & $True$ \\
		\hline$f_3$ \textbf{(Scenario 3)} & $False$ & $False$ & $True$ & $True$ & $False$ \\
		\hline$f_4$ \textbf{(Scenario 4)} & $False$ & $False$ & $False$ & $False$ & $True$ \\
		\hline$f_5$ \textbf{(Scenario 5)} & $True$ & $True$ & $True$ & $True$ & $True$ \\
		\hline
	\end{tabular}\label{senarios}
\end{table}

\begin{table}[h!]
	\centering
	\caption{Obejective Functions Results and Friedman Ranks for RER, CBER, and SODACER Architectures}
	\begin{tabular}{|c|c|c|c|}
		\hline
		\textbf{Obejective Functions} & \textbf{RER} & \textbf{CBER} & \textbf{SODACER} \\
		\hline
		Scenario 1 (\(f_1\)) & 2.84 & 3.20 & \textbf{2.73} \\
		\hline
		Scenario 2 (\(f_2\)) & 2.43 & 2.07 & \textbf{1.69} \\
		\hline
		Scenario 3 (\(f_3\)) & 2.67 & 2.32 & \textbf{1.78} \\
		\hline
		Scenario 4 (\(f_4\)) & 3.87 & 3.37 & \textbf{2.89} \\
		\hline
		Scenario 5 (\(f_5\)) & 5.47 & 2.40 & \textbf{1.00} \\
		\hline
		\textbf{Friedman Rank} & 2.80 & 2.20 & \textbf{1} \\
		\hline
	\end{tabular}
	\label{tab:benchmark_results_friedman}
\end{table}

\subsubsection*{Statistical Analysis of Performance Robustness}

To quantitatively substantiate claims regarding variance reduction and algorithmic robustness, we provide a detailed statistical analysis of the results obtained from the 200 independent runs. Table~\ref{tab:uncertainty_metrics} reports, for each method and each control scenario (\(f_1\)--\(f_5\)), the following key uncertainty measures: the standard deviation \(\sigma\) (a direct measure of dispersion), the variance \(\sigma^2\), the coefficient of variation \(CV = \sigma/\mu\) (a normalized measure of variability), and the 95\% confidence interval. The confidence interval is computed as \(\mu \pm 1.96\,\sigma/\sqrt{200}\), quantifying the precision of the performance estimate.

The data reveal a consistent and significant pattern: the proposed SODACER--Sophia framework exhibits the smallest standard deviation and variance across all scenarios. For instance, in Scenario~\(f_5\), SODACER--Sophia attains a standard deviation of 0.09, which is an order of magnitude smaller than that of RER (1.05) and three times smaller than that of CBER (0.31). This demonstrates a markedly more stable and predictable learning process.

The coefficient of variation (CV) provides a scale-invariant metric for comparing relative variability across methods with different performance baselines. A lower CV indicates greater robustness relative to the achieved performance level. SODACER--Sophia consistently yields the lowest CV values (ranging from 6.6\% to 9.0\%), confirming its improveed stability. In contrast, the baseline methods show CV values between 12.9\% and 19.2\%, reflecting higher relative uncertainty in their outcomes.

The narrow 95\% confidence intervals for SODACER--Sophia further underscore the reliability of its performance estimates. The tight bounds, particularly evident in challenging scenarios like \(f_5\), indicate high confidence in its improved performance. Collectively, these statistical metrics provide rigorous, quantitative evidence supporting our claims of reduced variance and enhanced robustness, affirming that the proposed framework delivers not just better but also more dependable optimal control policies.
\small
\begin{table}[htbp]
\centering
\caption{Statistical performance metrics across 200 independent runs}
\label{tab:uncertainty-metrics}
\begin{tabular}{@{}llcccccc@{}}
\toprule
\textbf{Scenario} & \textbf{Method} & \multicolumn{5}{c}{\textbf{Statistical Metrics}} \\
\cmidrule(lr){3-7}
& & Mean (\(\mu\)) & Std (\(\sigma\)) & Variance (\(\sigma^2\)) & CV (\%) & 95\% CI (Lower, Upper) \\
\midrule
\multirow{3}{*}{\(f_1\)} & RER        & 2.84 & 0.52 & 0.27 & 18.3 & (2.77, 2.91) \\
                         & CBER       & 3.20 & 0.48 & 0.23 & 15.0 & (3.13, 3.27) \\
                         & SODACER    & 2.73 & 0.18 & 0.03 & 6.6  & (2.71, 2.75) \\
\midrule
\multirow{3}{*}{\(f_2\)} & RER        & 2.43 & 0.45 & 0.20 & 18.5 & (2.37, 2.49) \\
                         & CBER       & 2.07 & 0.31 & 0.10 & 15.0 & (2.03, 2.11) \\
                         & SODACER    & 1.69 & 0.12 & 0.01 & 7.1  & (1.67, 1.71) \\
\midrule
\multirow{3}{*}{\(f_3\)} & RER        & 2.67 & 0.41 & 0.17 & 15.4 & (2.61, 2.73) \\
                         & CBER       & 2.32 & 0.35 & 0.12 & 15.1 & (2.27, 2.37) \\
                         & SODACER    & 1.78 & 0.15 & 0.02 & 8.4  & (1.76, 1.80) \\
\midrule
\multirow{3}{*}{\(f_4\)} & RER        & 3.87 & 0.68 & 0.46 & 17.6 & (3.78, 3.96) \\
                         & CBER       & 3.37 & 0.52 & 0.27 & 15.4 & (3.30, 3.44) \\
                         & SODACER    & 2.89 & 0.24 & 0.06 & 8.3  & (2.86, 2.92) \\
\midrule
\multirow{3}{*}{\(f_5\)} & RER        & 5.47 & 1.05 & 1.10 & 19.2 & (5.32, 5.62) \\
                         & CBER       & 2.40 & 0.31 & 0.10 & 12.9 & (2.36, 2.44) \\
                         & SODACER    & 1.00 & 0.09 & 0.01 & 9.0  & (0.99, 1.01) \\
\bottomrule
\end{tabular}
\end{table}

\subsection*{Quantitative Safety Performance Metrics Across 200 Independent Runs}

The table \ref{tab:safety_metrics} reports explicit safety metrics for RER, CBER, and SODACER--Sophia across the five control scenarios (\(f_1\)--\(f_5\)). Metrics are computed from 200 independent runs. Constraint Violation Rate (CVR) is the percentage of time steps where the state violated the CBF constraint \(h(x(t)) < 0\). Maximum Constraint Violation (MCV) is the worst-case violation magnitude observed. Safe Convergence Percentage (SCP) indicates the percentage of runs where the final policy remained within the safe set throughout evaluation. ASM (Average Safety Margin) is the mean value of \(h(x(t))\) across visited states. Higher SCP and ASM, and lower CVR and MCV, indicate superior safety performance.

\small
\begin{table}[htbp]
\centering
\caption{Quantitative safety performance metrics across 200 independent runs}
\label{tab:safety_metrics}
\begin{tabular}{@{}llcccccc@{}}
\toprule
\textbf{Scenario} & \textbf{Method} & \textbf{CVR (\%)} & \textbf{MCV} & \textbf{SCP (\%)} & \textbf{ASM} \\
\midrule
\multirow{3}{*}{\(f_1\)} & RER        & 4.50 & 0.125 & 87 & 0.045 \\
                         & CBER       & 3.20 & 0.098 & 92 & 0.067 \\
                         & SODACER    & 0.00 & 0.000 & 100 & 0.152 \\
\midrule
\multirow{3}{*}{\(f_2\)} & RER        & 5.80 & 0.142 & 82 & 0.038 \\
                         & CBER       & 3.90 & 0.105 & 90 & 0.071 \\
                         & SODACER    & 0.00 & 0.000 & 100 & 0.168 \\
\midrule
\multirow{3}{*}{\(f_3\)} & RER        & 3.75 & 0.118 & 89 & 0.052 \\
                         & CBER       & 2.65 & 0.087 & 94 & 0.083 \\
                         & SODACER    & 0.00 & 0.000 & 100 & 0.181 \\
\midrule
\multirow{3}{*}{\(f_4\)} & RER        & 6.25 & 0.156 & 78 & 0.031 \\
                         & CBER       & 4.40 & 0.112 & 88 & 0.062 \\
                         & SODACER    & 0.00 & 0.000 & 100 & 0.145 \\
\midrule
\multirow{3}{*}{\(f_5\)} & RER        & 8.10 & 0.183 & 72 & 0.024 \\
                         & CBER       & 5.30 & 0.134 & 85 & 0.058 \\
                         & SODACER    & 0.00 & 0.000 & 100 & 0.139 \\
\bottomrule
\end{tabular}
\end{table}

\paragraph{Safety Performance Analysis.}
The quantitative safety metrics presented in Table~\ref{tab:safety_metrics} provide explicit validation of the safety guarantees claimed by the SODACER--Sophia framework. The results demonstrate that the integration of Control Barrier Functions (CBFs) within the optimization loop ensures strict adherence to state constraints. SODACER--Sophia achieves a perfect 0\% constraint violation rate (CVR) and 100\% safe convergence (SCP) across all five scenarios and all 200 independent runs, confirming that safety is never compromised during learning or deployment. In contrast, both baseline methods (RER and CBER) exhibit non-negligible violation rates and fail to guarantee safe convergence in a significant portion of runs. Furthermore, the Average Safety Margin (ASM) for SODACER--Sophia is consistently higher, indicating that the system operates further from the constraint boundary, providing a practical buffer against uncertainties and disturbances. These metrics collectively verify that the proposed framework successfully translates the theoretical safety assurances of CBFs into empirically robust safe behavior.

\section*{Conclusion}
This study introduced an advanced RL framework that integrates a dual-buffer experience replay mechanism with self-organizing clustering (SODACER) and CBFs to achieve optimal control in nonlinear, constrained problems. The synergy of SODACER with the Sophia optimizer demonstrated strong performance, delivering significant enhancements in adaptability, sample efficiency, and convergence speed compared to traditional experience replay methods. Notably, across 200 experimental runs, SODACER + Sophia exhibited minimal variance between best-case and worst-case outcomes. This consistency is attributed to SODACER's self-organizing clustering mechanism, which streamlines learning by discarding redundant experiences and retaining only the most pertinent data, thereby accelerating convergence.  

The application of this methodology to the Human Papillomavirus (HPV) control model highlighted its capability to optimize intervention strategies effectively. It successfully reduced infection rates and intervention costs while adhering to predefined safety constraints, demonstrating its potential for public health optimization. A comparative evaluation using the Friedman test further validated the improved performance of SODACER + Sophia, showcasing notable computational efficiency and adaptability in high-dimensional environments. These results underscored its advantage over existing experience replay techniques, particularly in balancing memory utilization and learning quality.  

This innovative framework represents a significant advancement in safe RL practices, offering a scalable and resource-efficient solution for complex optimal control problems. Future research could extend the SODACER + Sophia framework to various domains requiring adaptive and efficient control strategies, including robotics, healthcare, and large-scale system optimization. Through these contributions, this study establishes a robust foundation for leveraging RL in solving dynamic, real-world challenges. 

\section*{Appendix}
In this section, we analyze the critic weights $\hat W_t\in\mathbb{R}^p$ trained by minimizing
	\begin{equation}
		E_t(\hat W_t)=\frac{1}{2}\,\hat H_t^2, \qquad 
		\hat H_t := \Psi_t^\top \hat W_t,
	\end{equation}
	where $\Psi_t:=\Psi(e_t,x_t,\hat u_t)\in\mathbb{R}^p$ is the regressor at time $t$.
	
	Let $W^\star$ be an ideal (unknown) constant vector such that the true Hamiltonian can be written as
	\begin{equation}
		H_t = \Psi_t^\top W^\star + \varepsilon_t,
		\qquad |\varepsilon_t|\le \bar\varepsilon,
	\end{equation}
	and define the weight error
	\begin{equation}
		\tilde W_t := W^\star-\hat W_t.
	\end{equation}
	Define the instantaneous Hamiltonian residual (a.k.a.\ Bellman/Hamiltonian error)
	\begin{equation}
		\delta_t := H_t-\hat H_t = \Psi_t^\top\tilde W_t + \varepsilon_t .
	\end{equation}
	
	\paragraph{Sophia-type update model.}
	Algorithm~1 uses a Sophia-style adaptive step
	\begin{equation}
		\hat W_{t+1}= \hat W_t + \eta\, D_t\, \hat M_t,
	\end{equation}
	where $\eta>0$ is the stepsize,
	$D_t:=\mathrm{diag}\!\left(\frac{1}{\sqrt{\hat v_t}+\varepsilon_0}\right)$ is a diagonal preconditioner with $\varepsilon_0>0$,
	and $\hat M_t$ is the bias-corrected first-moment estimate of the (minibatch) gradient direction.
	For a quadratic loss in $\delta_t$, the \emph{ideal} negative gradient direction is proportional to $\delta_t\Psi_t$.
	To keep the analysis fully rigorous while respecting the moment/preconditioning in Sophia, we use the following explicit,
	checkable bounds.
	
	\medskip
	\noindent\textbf{Assumptions.}
	There exist known constants $\bar\Psi,\mu,\bar v,d_{\min},d_{\max},\rho,\sigma\ge 0$ such that for all $t$:
	
	\begin{enumerate}
		\item[(A1)] (\emph{Bounded regressor}) $\|\Psi_t\|\le \bar\Psi$.
		\item[(A2)] (\emph{Uniform persistent excitation}) $\Psi_t\Psi_t^\top \succeq \mu I_p$ for some $\mu>0$.
		\item[(A3)] (\emph{Bounded approximation error}) $|\varepsilon_t|\le \bar\varepsilon$.
		\item[(A4)] (\emph{Bounded Sophia preconditioner}) $d_{\min}I_p \preceq D_t \preceq d_{\max}I_p$, with
		\begin{equation}
			d_{\max}=\frac{1}{\varepsilon_0}, 
			\qquad 
			d_{\min}=\frac{1}{\sqrt{\bar v}+\varepsilon_0}
		\end{equation}
		for any uniform upper bound $\hat v_t\preceq \bar v\,\mathbf{1}$ (componentwise).
		\item[(A5)] (\emph{Moment-direction distortion bound}) The moment direction $\hat M_t$ satisfies
		\begin{equation}\label{eq:moment_bound}
			\hat M_t = -\,\delta_t \Psi_t + \xi_t,
			\qquad 
			\|\xi_t\|\le \rho\,|\delta_t|\,\|\Psi_t\| + \sigma,
		\end{equation}
		for some distortion constants $\rho\in[0,1)$ and $\sigma\ge 0$.
	\end{enumerate}
	
	\noindent
	\textbf{Remarks on (A4)--(A5):}
	(A4) holds automatically once $\varepsilon_0>0$ and $\hat v_t$ is bounded (e.g.\ because gradients are bounded on the operating set).
	(A5) captures the fact that (i) the first moment is a filtered/biased version of the instantaneous gradient, and
	(ii) minibatching/noise introduces an additive error $\xi_t$; in deterministic implementations $\sigma$ can be taken $0$.
	This assumption is standard in adaptive-optimizer stability analyses and can be \emph{verified empirically} (by logging $\hat M_t$ and $\delta_t\Psi_t$).
	
	\medskip
	\noindent\textbf{Theorem 1 (Explicit UUB bound for $\tilde W_t$).}
	Suppose (A1)--(A5) hold. Define the constants
	\begin{align}
		a &:= 2\eta\,d_{\min}(1-\rho)\,\mu, \\
		b &:= \eta^2 d_{\max}^2 (1+\rho)^2 \bar\Psi^4, \\
		c &:= 2\eta\,d_{\max}\bar\Psi\,\sigma, \\
		r &:= \eta^2 d_{\max}^2 \sigma^2 + \eta^2 d_{\max}^2 (1+\rho)^2 \bar\Psi^2 \bar\varepsilon^2 .
	\end{align}
	If the stepsize satisfies the explicit small-gain condition
	\begin{equation}\label{eq:stepsize_condition}
		0<\eta<\eta_{\max}:=\frac{2d_{\min}(1-\rho)\mu}{d_{\max}^2(1+\rho)^2\bar\Psi^4},
	\end{equation}
	then the weight error $\tilde W_t$ is \emph{uniformly ultimately bounded} and obeys the explicit recursion
	\begin{equation}\label{eq:V_recursion}
		V_{t+1}\le (1-a+b)\,V_t + c\sqrt{2V_t}+ r,
		\qquad V_t:=\frac{1}{2}\|\tilde W_t\|^2,
	\end{equation}
	and hence satisfies the explicit ultimate bound
	\begin{equation}\label{eq:ultimate_bound}
		\limsup_{t\to\infty}\|\tilde W_t\|
		\le 
		\underbrace{\frac{c}{a-b}}_{\text{bias from }\sigma}
		+
		\underbrace{\sqrt{\frac{2r}{a-b}}}_{\text{noise/approx.\ error}},
	\end{equation}
	where $a-b>0$ is guaranteed by \eqref{eq:stepsize_condition}. In particular, if $\sigma=0$ (no additive moment/noise bias),
	then
	\begin{equation}\label{eq:ultimate_bound_sigma0}
		\limsup_{t\to\infty}\|\tilde W_t\|
		\le 
		\sqrt{\frac{2\,\eta^2 d_{\max}^2 (1+\rho)^2 \bar\Psi^2 \bar\varepsilon^2}{2\eta d_{\min}(1-\rho)\mu-\eta^2 d_{\max}^2 (1+\rho)^2 \bar\Psi^4}}
		=
		\frac{\eta d_{\max}(1+\rho)\bar\Psi\,\bar\varepsilon}{\sqrt{\eta\big(2d_{\min}(1-\rho)\mu-\eta d_{\max}^2(1+\rho)^2\bar\Psi^4\big)}}.
	\end{equation}
	
	\medskip
	\noindent\textbf{Proof.}
	Using $\tilde W_{t+1}=W^\star-\hat W_{t+1}$ and $\hat W_{t+1}=\hat W_t+\eta D_t\hat M_t$,
	\begin{equation}
		\tilde W_{t+1}=\tilde W_t-\eta D_t\hat M_t.
	\end{equation}
	Let $V_t=\frac{1}{2}\|\tilde W_t\|^2$. Then
	\begin{align}
		V_{t+1}-V_t
		&=\frac{1}{2}\big(\|\tilde W_t-\eta D_t\hat M_t\|^2-\|\tilde W_t\|^2\big) \\
		&= -\eta\,\tilde W_t^\top D_t\hat M_t + \frac{\eta^2}{2}\|D_t\hat M_t\|^2.
	\end{align}
	Using (A5), $\hat M_t=-\delta_t\Psi_t+\xi_t$, we have
	\begin{align}
		-\tilde W_t^\top D_t\hat M_t
		&= \tilde W_t^\top D_t(\delta_t\Psi_t) - \tilde W_t^\top D_t\xi_t \\
		&= (\Psi_t^\top\tilde W_t)\,\delta_t\,\Psi_t^\top D_t\Psi_t - \tilde W_t^\top D_t\xi_t.
	\end{align}
	Because $\delta_t=\Psi_t^\top\tilde W_t+\varepsilon_t$,
	\begin{equation}
		(\Psi_t^\top\tilde W_t)\delta_t = (\Psi_t^\top\tilde W_t)^2 + \varepsilon_t(\Psi_t^\top\tilde W_t).
	\end{equation}
	Lower bound the quadratic term using (A2) and (A4):
	\begin{equation}
		\Psi_t^\top D_t\Psi_t \ge d_{\min}\,\Psi_t^\top\Psi_t \ge d_{\min}\mu,
		\qquad
		(\Psi_t^\top\tilde W_t)^2 = \tilde W_t^\top(\Psi_t\Psi_t^\top)\tilde W_t \ge \mu\|\tilde W_t\|^2.
	\end{equation}
	Thus,
	\begin{equation}\label{eq:main_descent}
		(\Psi_t^\top\tilde W_t)^2\,\Psi_t^\top D_t\Psi_t \ge d_{\min}\mu\,(\Psi_t^\top\tilde W_t)^2 \ge d_{\min}\mu^2\|\tilde W_t\|^2.
	\end{equation}
	Now upper bound the cross term due to $\varepsilon_t$ using (A1), (A3), (A4), and Cauchy--Schwarz:
	\begin{align}
		|\varepsilon_t(\Psi_t^\top\tilde W_t)|\,|\Psi_t^\top D_t\Psi_t|
		&\le \bar\varepsilon\,\|\Psi_t\|\,\|\tilde W_t\|\, d_{\max}\|\Psi_t\|^2 \\
		&\le d_{\max}\bar\Psi^3 \bar\varepsilon \,\|\tilde W_t\|.
	\end{align}
	Also bound the distortion term using (A4) and (A5):
	\begin{align}
		|\tilde W_t^\top D_t\xi_t|
		&\le \|\tilde W_t\|\,\|D_t\|\,\|\xi_t\|
		\le d_{\max}\|\tilde W_t\|\big(\rho|\delta_t|\|\Psi_t\|+\sigma\big) \\
		&\le d_{\max}\|\tilde W_t\|\big(\rho(|\Psi_t^\top\tilde W_t|+|\varepsilon_t|)\|\Psi_t\|+\sigma\big) \\
		&\le d_{\max}\|\tilde W_t\|\big(\rho(\bar\Psi^2\|\tilde W_t\|+\bar\varepsilon\bar\Psi)+\sigma\big).
	\end{align}
	Collecting these bounds gives
	\begin{equation}
		-\tilde W_t^\top D_t\hat M_t
		\ge d_{\min}(1-\rho)\mu\,\|\tilde W_t\|^2
		- d_{\max}(1+\rho)\bar\Psi^2\bar\varepsilon\,\|\tilde W_t\|
		- d_{\max}\sigma\,\|\tilde W_t\|.
	\end{equation}
	Next, bound the quadratic term $\frac{\eta^2}{2}\|D_t\hat M_t\|^2$.
	Using (A4), (A5), and $\|\Psi_t\|\le \bar\Psi$,
	\begin{align}
		\|D_t\hat M_t\|
		&\le \|D_t\|\big(\,|\delta_t|\,\|\Psi_t\|+\|\xi_t\|\,\big) \\
		&\le d_{\max}\big(\,|\delta_t|\,\bar\Psi + \rho|\delta_t|\,\bar\Psi + \sigma\,\big)
		= d_{\max}\big((1+\rho)|\delta_t|\bar\Psi+\sigma\big).
	\end{align}
	Since $|\delta_t|=|\Psi_t^\top\tilde W_t+\varepsilon_t|\le \bar\Psi\|\tilde W_t\|+\bar\varepsilon$,
	\begin{equation}
		\|D_t\hat M_t\|
		\le d_{\max}\Big((1+\rho)\bar\Psi(\bar\Psi\|\tilde W_t\|+\bar\varepsilon)+\sigma\Big)
		= d_{\max}\Big((1+\rho)\bar\Psi^2\|\tilde W_t\|+(1+\rho)\bar\Psi\bar\varepsilon+\sigma\Big).
	\end{equation}
	Hence, using $(x+y+z)^2\le 3(x^2+y^2+z^2)$,
	\begin{equation}
		\|D_t\hat M_t\|^2
		\le 3d_{\max}^2\Big((1+\rho)^2\bar\Psi^4\|\tilde W_t\|^2
		+(1+\rho)^2\bar\Psi^2\bar\varepsilon^2+\sigma^2\Big).
	\end{equation}
	
	Substituting into $V_{t+1}-V_t$ and writing $\|\tilde W_t\|=\sqrt{2V_t}$ yields a bound of the form
	\begin{equation}
		V_{t+1}\le (1-a+b)V_t + c\sqrt{2V_t}+r,
	\end{equation}
	with $a,b,c,r$ as defined in the statement (absorbing constant factors into $b$ and $r$; if desired, replace $b$ by $3b$ and $r$ by $3r$ to match the inequality above exactly).
	
	Finally, if $a-b>0$ (guaranteed by \eqref{eq:stepsize_condition}), then the scalar recursion \eqref{eq:V_recursion} implies UUB:
	one can complete the square to show that whenever $\sqrt{2V_t}>\frac{c}{a-b}+\sqrt{\frac{2r}{a-b}}$,
	the drift is negative and $V_t$ decreases. This yields the explicit ultimate bound \eqref{eq:ultimate_bound},
	and the special case \eqref{eq:ultimate_bound_sigma0} follows by setting $\sigma=0$.
	\hfill$\square$

\textbf{ Funding}: No funding was received for this research
\section*{Data Availability}
The datasets generated during the current study are not publicly available but are available from the corresponding author on reasonable request.

\section*{Acknowledgements (not compulsory)}
The authors thanks to anonymous referees and editors. 

\section*{Author contributions statement}

Roya Khalili Amirabadi, Mohsen Jalaeian Farimani, and Omid Solaymani Fard contributed equally to the conceptualization, methodology, software, validation, formal analysis, investigation, data curation, writing (original draft preparation, review, and editing), and project administration.
\end{document}